\documentclass[lettersize,journal,10pt]{IEEEtran}
\usepackage{amsmath,amsfonts}
\usepackage{algpseudocode}
\usepackage{algorithm}
\usepackage{amsmath,amsfonts}
\usepackage{algpseudocode}
\usepackage{algorithm}
\usepackage{array}
\usepackage[caption=false,font=normalsize,labelfont=sf,textfont=sf]{subfig}
\usepackage{textcomp}
\usepackage{stfloats}
\usepackage{url}
\usepackage{verbatim}
\usepackage{color,soul}
\usepackage{graphicx}
\usepackage{hyperref}
\usepackage{cite}
\usepackage{makecell}
\usepackage[nolist]{acronym}
\usepackage{tikz}
\usepackage[utf8]{inputenc}
\hypersetup{draft}
\usepackage{titlesec}

\begin{acronym}
    \acro{MIMO}{multiple-input, multiple-output}
    \acro{AWGN}{additive white Gaussian noise}
    \acro{KPI}{key performance indicator}
      \acro{KPIs}{key performance indicators}  
    \acro{D2D}{device-to-device}
    \acro{ISAC}{integrated sensing and communication}
    \acro{DFRC}{dual-functional radar and communication}
    \acro{AoA}{angle of arrival}
    \acro{AoD}{angle of departure}
    \acro{AoAs}{angles of arrival}
    \acro{ToA}{time of arrival}
    \acro{EVM}{error vector magnitude}
    \acro{CPI}{coherent processing interval}
    \acro{eMBB}{enhanced mobile broadband}
    \acro{URLLC}{ultra-reliable low latency communications}
    \acro{mMTC}{massive machine type communications}
	\acro{QCQP}{quadratic constrained quadratic programming}
	\acro{BnB}{branch and bound}
	\acro{SER}{symbol error rate}
	\acro{LFM}{linear frequency modulation}
	\acro{BS}{base station}
	\acro{UE}{user equipment}
	\acro{DL}{downlink}
	\acro{UL}{uplink}
	\acro{CS}{compressed sensing}
	\acro{PR}{passive radar}
	\acro{LMS}{least mean squares}
	\acro{NLMS}{normalized least mean squares}
	\acro{RIS}{reconfigurable intelligent surface}
	\acro{RISs}{reconfigurable intelligent surfaces}
	\acro{IRS}{intelligent reflecting surface}
	\acro{LISA}{large intelligent surface/antennas}
	\acro{OFDM}{orthogonal frequency-division multiplexing}
	\acro{EM}{electromagnetic}
	\acro{mmWave}{milli-meter wave}
	\acro{MSE}{mean squared error}
		\acro{XR}{extended reality}
	\acro{SNR}{signal-to-noise ratio}
	\acro{SRP}{successful recovery probability}
	\acro{CDF}{cumulative distribution function}
	\acro{ULA}{uniform linear array}
	\acro{RSS}{received signal strength}
	\acro{SNDR}{signal-to-noise-and-distortion ratio}
	\acro{DPI}{direct path interference}
	\acro{RPI}{reflected path interference}
	\acro{DE}{direct echo}
	\acro{RE}{RIS-resulted echo}
	\acro{PI}{path interference}
	\acro{ADC}{analog-to-digital converter}
	\acro{RF}{radio frequency}
	\acro{BCD}{block coordinate descent}
	\acro{BCCD}{block cyclic coordinate descent}
	\acro{RCG}{Riemannian conjugate gradient}
	\acro{SDP}{semi-definite program}
	\acro{EVD}{eigenvalue decomposition}
		\acro{RCS}{radar cross-section}
		\acro{STAR}{simultaneous transmitting and reflecting}
		\acro{LNA}{low noise amplifier}
  \acro{NOMA}{non-orthogonal multiple access}
  \acro{FM}{frequency modulation}
  \acro{SINR}{signal to interference plus noise ratio}
  \acro{V2X}{Vehicle-to-Everything}
  \acro{LoS}{line-of-sight}
  \acro{NLoS}{non-line-of-sight}
  \acro{CPU}{central processing unit}
  \acro{DoF}{degrees-of-freedom}
  \acro{CRB}{Cram\'er-Rao bound}
  \acro{CSI}{channel state information}
\end{acronym}

\DeclareMathOperator{\dB}{dB}

\DeclareMathOperator{\Real}{Re}

\DeclareMathOperator{\dBm}{dBm}
\DeclareMathOperator{\dBi}{dBi}
\DeclareMathOperator{\meters}{m}
\DeclareMathOperator{\GHz}{GHz}
\DeclareMathOperator{\dBmpHz}{dBm/Hz}

\DeclareMathOperator{\grad}{grad}
\DeclareMathOperator{\Retr}{Retr}

\newcommand\myeqa{\mathrel{\overset{\makebox[0pt]{\mbox{\normalfont\tiny\sffamily (a)}}}{=}}}
\newcommand\myeqb{\mathrel{\overset{\makebox[0pt]{\mbox{\normalfont\tiny\sffamily (b)}}}{=}}}
\newcommand\myeqc{\mathrel{\overset{\makebox[0pt]{\mbox{\normalfont\tiny\sffamily (c)}}}{=}}}
\newcommand\myeqd{\mathrel{\overset{\makebox[0pt]{\mbox{\normalfont\tiny\sffamily (d)}}}{=}}}
\newcommand\myeqe{\mathrel{\overset{\makebox[0pt]{\mbox{\normalfont\tiny\sffamily (e)}}}{=}}}
\newcommand\myeqf{\mathrel{\overset{\makebox[0pt]{\mbox{\normalfont\tiny\sffamily (f)}}}{=}}}

\usepackage{xcolor,cite,etoolbox}
\makeatletter 
\pretocmd\@bibitem{\color{black}\csname keycolor#1\endcsname}{}{\fail}
\newcommand\citecolor[1]{\@namedef{keycolor#1}{\color{blue}}}
\makeatother

\titlespacing{\section}{0pt}{*0.2}{*0.1}
\begin{document}
\bstctlcite{IEEEexample:BSTcontrol}
\title{Low Dynamic Range for RIS-aided Bistatic Integrated Sensing and Communication}
\author{Ahmad Bazzi and Marwa Chafii 
\thanks{

Ahmad Bazzi and Marwa Chafii are with the Engineering Division, New York University (NYU) Abu Dhabi, 129188, UAE and NYU WIRELESS,
NYU Tandon School of Engineering, Brooklyn, 11201, NY, USA
(email: \href{ahmad.bazzi@nyu.edu}{ahmad.bazzi@nyu.edu}, \href{marwa.chafii@nyu.edu}{marwa.chafii@nyu.edu}).
.}
\thanks{Manuscript received xxx}}
\markboth{accepted in \textit{IEEE Journal on Selected Areas in Communications}}%
{Shell \MakeLowercase{\textit{et al.}}: A Sample Article Using IEEEtran.cls for IEEE Journals}


\maketitle

\begin{abstract}
The following paper presents a reconfigurable intelligent surface (RIS)-aided integrated sensing and communication (ISAC) system model scenario, where a base station communicates with a user, and a bi-static sensing unit, i.e. the passive radar (PR), senses targets using downlink signals. Given that the RIS aids with communication and sensing tasks, this paper introduces new interfering paths that can overwhelm the PR with unnecessarily high power, namely the path interference (PI), \textit{which is itself a combination of two interfering paths, the direct path interference (DPI) and the reflected path interference (RPI)}.
For this, we formulate an optimization framework that allows the system to carry on with its ISAC tasks, through analog space-time beamforming at the sensing unit, in collaboration with RIS phase shift and statistical transmit covariance matrix optimization, while minimizing the PI power.
As the proposed optimization problem is non-convex, we tailor a block-cyclic coordinate descent (BCCD) method to decouple the non-convex sub-problem from the convex one. 
A Riemannian conjugate gradient method is devised to generate the RIS and PR space-time beamforming phase shifts per BCCD iteration, while the convex sub-problem is solved via off-the-shelf solvers.
Simulation results demonstrate the effectiveness of the proposed solver when compared with benchmarking ones.
\end{abstract}
\begin{IEEEkeywords}
Integrated Sensing and Communication, ISAC, Reconfigurable Intelligent Surfaces, RIS, 6G
\end{IEEEkeywords}

\section{Introduction}
\label{sec:introduction}
Research in the realm of $6$G anticipates future services and applications to address diverse market demands and disruptive technologies \cite{10041914}. $6$G's objectives include supporting a range of services such as blockchain, haptic telemedicine, VR/AR remote services, holographic teleportation, and \ac{XR}. This transcends conventional connectivity, incorporating sensing, computing, and leveraging environmental data for artificial intelligence and machine learning. However, bandwidth-intensive applications require a substantial capacity increase, expected to surge by a factor of $10^3$ by $2030$ \cite{saad2019vision}. Crucial roles are attributed to enabling technologies like sub-$6$ GHz, \ac{mmWave} frequencies, and \ac{RIS}.
$6$G is also anticipated to support \ac{ISAC} applications.
Indeed, radar sensing and wireless communication share numerous similarities in signal processing and system design \cite{9585321}. 
\ac{ISAC} holds promise for groundbreaking applications, particularly in the automotive sector \cite{9830717}, the Internet of things, and robotics. \ac{ISAC} collaboration and convergence are crucial features that can pave the way for innovations such as \ac{V2X} communication \cite{9606831}, leveraging high-rate communications and high-precision localization capabilities.
\ac{ISAC} has been explored across various scenarios and challenges, including \ac{NOMA} \cite{9668964}, holographic communications \cite{9724245}, waveform design \cite{bazzi2022integrated}, security \cite{10373185}, and beamforming design \cite{10018908}.

\acp{RIS} \cite{10024150,10053657,10143420,10124870,10155675}, comprised of adaptable reflecting elements, have garnered significant attention for their potential to enhance the capacity and coverage of wireless sensor networks \cite{renzo2019smart}. These structures manipulate incident signals in the wireless propagation environment, thereby altering wireless channels between network nodes. \acp{RIS} provide a low-energy consumption solution, enabling signal amplification without the need for power amplifiers. Their cost-effectiveness could lead to large-scale deployment, covering entire building walls \cite{han2022localization}. The appropriately designed phase shifts from each reflecting element collectively combine reflected signals, commonly referred to as \ac{LISA} or \ac{IRS} in the literature. \acp{RIS} find applications in various wireless communication aspects, including beamforming, security, \ac{OFDM}, and millimeter-wave channel estimation \cite{lin2020reconfigurable, xu2022ris, zhou2022channel}. Their deployment enhances spectral and energy efficiency, supports environmentally friendly deployment, and contributes to secure wireless communication \cite{huang2019reconfigurable, liaskos2018using}. On the other hand, \ac{PR}, a radar design that detection and localization of targets without emitting controlled radar data, has been extensively studied \cite{dawidowicz2012detection, strinati2021wireless}. Moreover, \ac{RIS} is anticipated to play a pivotal role in advanced localization and sensing through \ac{PR}. 
In addition, \ac{PR} offers cost-effective procurement, operation, and maintenance, as well as covert operation. However, it encounters challenges in target detection due to the absence of transmitted waveform information. 
Nonetheless, it remains protocol-agnostic and operates without active component assistance.

To begin, forthcoming wireless networks designed to accommodate \ac{ISAC} are anticipated to confront challenges related to paths undergoing single, double, and triple pathloss, particularly with the recent advancements in \ac{RIS}. 
 Despite these developments, little attention has been given to dynamic range considerations for \ac{ISAC}, creating a significant gap in addressing this critical issue. 
 Specifically, the dynamic range, representing the ratio between the highest and lowest power levels of signal components, determines the number of bits required by a receiver's \ac{ADC} to minimize quantization errors for accurate signal processing. 
However, simply increasing the number of bits results in greater power consumption.
 Hence, it is imperative to precisely identify the relevant signal components to economize the dynamic range efficiently.
 
It is important to note that there is ongoing research works on \ac{RIS} \cite{9424177} for \ac{ISAC} systems to advance the road towards $6$G \cite{Wu20241}.
The work in \cite{Ye20241} designs \ac{RIS}-aided \ac{ISAC} beamforming for optimizing spectrum utilization via lightweight unsupervised learning-based methods that can achieve high spectrum efficiency. 
Also, \cite{Sankar20244017} considers beamforming for \ac{RIS}-\ac{ISAC} systems to  precode both communication symbols and radar waveforms.
Additionally, \cite{Liu20231} explores an \ac{RIS}-assisted \ac{ISAC} system where a multi-antenna \ac{BS} handles both multi-user communications and radar sensing, focusing on optimizing target detection and parameter estimation. Two non-convex optimization problems are addressed to maximize the sum-rate under \ac{SNR} and \ac{CRB} constraints, with algorithms developed to solve the optimization problems.
Furthermore, \cite{Chen20242696}  presents an \ac{ISAC} system utilizing \ac{RIS}, with  simultaneous beam training and target sensing that differentiates between \ac{RIS} and target echoes, with low-complexity positioning and array orientation estimation methods to facilitate beam alignment.
In addition, \cite{10571090} derives sensing algorithms for joint \ac{AoA} and \ac{ToA} estimation in a communication-centric \ac{RIS} assisted bistatic scenario incorporating a \ac{PR}.
The work in \cite{Xu20242232} addresses the challenge of limited \ac{DoF} in \ac{ISAC} systems by leveraging \ac{RIS} to enhance beamforming capabilities. For this, two optimization techniques are proposed: an alternating optimization algorithm combining semidefinite relaxation and one-dimension iteration to maximize radar mutual information, and another to optimize weighted \ac{ISAC} performance.
In terms of co-existence, \cite{He20222131} proposes a double-\ac{RIS}-assisted \ac{ISAC} system to enhance communication signals and suppress mutual interference in communication-radar coexistence scenarios. More specifically, \cite{He20222131} introduces a penalty dual decomposition-based algorithm for joint beamforming optimization, addressing both high and low radar power cases.
From a security perspective, \cite{Wei20241} focuses on the security challenges arising in \ac{ISAC} systems by optimizing a \ac{STAR}-\ac{RIS} \cite{9570143} assisted setup, which maximizes the sum secrecy rate while ensuring the necessary beampattern gain for target sensing, through proper optimization techniques. In the context of \ac{STAR}-\ac{RIS} for \ac{ISAC}, \cite{Li20241} studies a structured transmission structure for the \ac{ISAC} schemes in high mobility scenarios.

The focus of this paper is on a situation where a \ac{PR} is exposed to a signal-of-opportunity arising from communication signals transmitted by the \ac{BS}, and further reflected off-targets, ultimately reaching the \ac{PR}.
As the bi-static configuration is widely utilized in radar applications, it presents a notable issue that could potentially elevate costs for future \ac{ISAC} systems.
The \ac{DPI} is a common problem within the bi-static radar literature \cite{9968093}, and its suppression is crucial for practical radar operations \cite{9968093},\cite{jkedrzejewski2024experimental}.
With the introduction of \ac{RIS}, the power imbalances stemming from single, double, and triple path-loss will overshadow the \ac{PR} with unnecessary high-power differentials. 
This is primarily attributed to the \ac{BS}$\rightarrow$\ac{PR} and \ac{BS}$\rightarrow$\ac{RIS}$\rightarrow$\ac{PR} paths in contrast to other sensing paths involving the target.
Analog domain suppression is one direction that can be employed to mitigate the \ac{DPI} in bi-static and passive radar. Indeed, if we were lucky enough to choose an \ac{ADC} with infinite resolution, delegating part of this cancellation to analog would not be necessary..
So, analog domain processing, including analog beamforming, can help in reducing the dynamic range before digital conversion when low-cost, low-power, or low-resolution \acp{ADC} are integrated, so as not to degrade the received signal purity.
Moreover, analog domain processing is simple as the components involved typically consist of couplers and phase compensators \cite{6832471}; yet providing substantial suppression capabilities which can vary from $30$ to $70\dB$, depending on the implementation complexity \cite{6832464}.
In this context, a \ac{DPI} suppression method was implemented in the analog domain before the digital conversion was proposed in \cite{8376381} for \ac{FM} based passive radars.
Related suppression techniques may include methods such as null-steering and physical shielding, which are only moderately effective and can limit the system's feasibility \cite{7168278}. 
In this paper, and inspired by the reasoning of delegating interference cancellation to the analog domain, we allow the \ac{PR} to incorporate space-time beamforming, by adjusting phase shifts across space and time in order to pragmatically mitigate any potential interfering tasks, while allowing the \ac{PR} to carry on its sensing task.

This work focuses on an \ac{ISAC} system model in the presence of a receiving sensing unit, i.e. the \ac{PR} and a \ac{RIS} where the goal is to optimize for communication \acp{SNR} and radar \acp{SNDR} while maintaining a low \ac{PI} at the \ac{PR}.
To that purpose, we have summarized our contributions as follows.
\begin{itemize}
	\item \textbf{\ac{RIS}-aided \ac{ISAC} system with  \ac{PR}}. We model a \ac{RIS}-aided \ac{ISAC} system model, where the \ac{BS} along with the \ac{RIS} can together help the \ac{PR}, reduce the dynamic range due to high \ac{PI}, in conjunction with analog beamforming at the \ac{PR}, while performing \ac{ISAC} tasks.
	\item \textbf{Figure of merit and motivation} We extend the definition of \ac{DPI}, which is a known interference path in bistatic models, to the more general \ac{PI} which arises due to the presence of a \ac{RIS} in the scene. Essentially, the \ac{PI} is the sum of the \ac{DPI} and the \ac{RPI}, where the later is due to the \ac{BS}$\rightarrow$\ac{RIS}$\rightarrow$\ac{PR} path. We then provide context as to why this figure-of-merit is important, and the reverberations it has on \textit{flooding the sensing architecture with a non-reasonably high dynamic range}. 
	\item \textbf{\ac{RIS}-\ac{ISAC} optimization framework}. We propose an optimization framework tailored for minimizing the resulting high dynamic range while maintaining given \ac{ISAC} performance in terms of communication \ac{SNR}, as well as radar \ac{SNDR}. In particular, the \ac{BS} optimizes its statistical covariance matrix, along with the \ac{RIS} phase shifts and the \ac{PR} analog phase shifts to jointly minimize the strong \ac{PI} component, whilst maintaining given radar \ac{SNDR} and communication \acp{SNR} levels.
	\item \textbf{Solution via block cyclic coordinate descent}. As the proposed \ac{RIS}-\ac{ISAC} optimization problem is non-convex and highly non-linear, we design a \ac{BCCD} algorithm that decomposes the problem into two sub-problems, one of which contains is non-linear and non-convex, and solved via an \ac{RCG} method, and the other is an \ac{SDP}, hence convex. The \ac{BCCD} then iterates between the two problems and converges to a stable solution providing the statistical transmit covariance matrix, as well as the \ac{RIS} phase shifts and \ac{PR} analog phase shifts. 
	\item \textbf{Computational Complexity Analysis}. We  present an analysis of the computational complexity associated with the solutions of the proposed optimization problems for low dynamic range bistatic \ac{ISAC} in terms of \textit{"big-oh"} complexity.
	\item \textbf{Extensive simulation results}. We present extensive simulation results that highlight the superiority, as well as the potential of the proposed design and algorithm with respect to many benchmarks, in terms of achievable dynamic ranges, radar \ac{SNDR} and communication \ac{SNR}.   
\end{itemize}
Moreover, we unveil some important insights, i.e.
\begin{itemize}
	\item Increasing the number of transmit antennas effectively reduces the dynamic range required at \ac{PR}, which was observed across all benchmarks and the proposed method. Specifically, in scenarios with high $M_t$, doubling the number of antennas tends to decrease the dynamic range by approximately $5\dB$. Moreover, the quantity of antennas at the PR enhances analog beamforming capabilities, further reducing the dynamic range needed. For instance, transitioning from $1$ to $16$ antennas presents a near $10\dB$ dynamic range reduction, with an additional $10\dB$ decrease when quadrupling from $16$ to $64$ antennas using the proposed method. Remarkably, the proposed method achieves a dramatic reduction in dynamic range by up to $50\dB$ compared to setups without a \ac{RIS}, highlighting the method's efficiency in integrating \ac{PI} management directly into the optimization process.
	\item  The proposed method reveals a significant advantage in achieving high communication SNR at substantially lower dynamic ranges compared to benchmark methods, indicating its efficiency in optimizing system performance with limited dynamic range. For instance, to achieve a communication \ac{SNR} of $30\dB$, the proposed approach requires only a $23\dB$ dynamic range with only a single antenna at the \ac{PR}, whereas benchmarks necessitate over $100\dB$. Secondly, a counter-intuitive observation emerges where increasing the number of antennas $M$ allows for enhanced communication \ac{SNR} at a fixed dynamic range. For example, a dynamic range of $20\dB$ achieves a $26\dB$ communication \ac{SNR} with one antenna but can reach $35\dB$ \ac{SNR} with $16$ antennas and exceed $40\dB$ with $64$ antennas.
	\item We observe that prioritizing communication tasks over sensing tasks can significantly reduce the \ac{PI} power level for fixed resources (i.e. \ac{RIS} size and antennas at the \ac{PR}). Specifically, simulations reveal that increasing the communication priority leads to a marked decrease in \ac{PI} power levels, particularly with the proposed method demonstrating the most substantial reduction, highlighting the effectiveness of communication prioritization in mitigating \ac{PI} in \ac{RIS}-\ac{ISAC} systems.
\end{itemize}

The following paper is organized as follows: 
Section \ref{sec:system-model} presents the \ac{ISAC} system model for the underlying problem in the presence of a \ac{RIS}.
Section \ref{sec:motivation-of-dynamic-range} describes the main motivation, in addition to the importance of addressing dynamic range in our \ac{ISAC} \ac{RIS}-aided problem.
Moreover, Section \ref{sec:dynamic-range-minimization} formulates an \ac{ISAC} optimization problem aiming at optimizing the dynamic range at the sensing unit, i.e. the \ac{PR}. In addition, we introduce an efficient solver to tackle the non-convex problem at hand.
Section \ref{sec:simulation-results} presents our simulation results.
Section \ref{sec:open-challenges} discusses open challenges for future \ac{RIS}-aided \ac{ISAC} bi-static systems.
Section \ref{sec:conclusion} concludes the paper.

\textbf{Notation}: Upper-case and lower-case boldface letters denote matrices and vectors, respectively. $(.)^T$, $(.)^*$ and $(.)^H$ represent the transpose, the conjugate and the transpose-conjugate operators. 
The set of all complex-valued $N \times M$ matrices is $\mathbb{C}^{N \times M}$. 
The trace is $\operatorname{tr}()$. 
The $N \times N $ identity matrix is $\pmb{I}_N$.
We index the $(i,j)^{th}$ element of $\pmb{A}$ as $\pmb{A}(i,j)$. 
The Kronecker product is $\otimes$.
The Hadamard product is $\odot$.
The $\operatorname{diag}()$ operator applied onto a vector $\pmb{x}$, i.e. $\operatorname{diag}(\pmb{x})$ returns a diagonal matrix consisting of entries of $\pmb{x}$.
$[\pmb{x}]_m$ returns the $m^{th}$ element of vector $\pmb{x}$.The statement $\pmb{A} \succeq \pmb{B}$ means that $\pmb{A} - \pmb{B}$ is a positive semi-definite matrix.
All other notations are defined within the manuscript's text.

\section{System Model}
\label{sec:system-model}
\begin{figure}[!t]
\input{Actions/figure-system-model.tex}
\label{fig_1}
\end{figure}
Consider an \ac{ISAC} system comprised of a target of interest, a multi-antenna communication user, and a communication \ac{BS}, an \ac{RIS}, and a \ac{PR}. 
The \ac{BS} is equipped with an antenna array composed of $M_t$ elements,
whereas the communication user is equipped with $M_r$ antennas.
Also, assume that a \ac{RIS} containing $N$ reflecting elements is present. 
In addition, the \ac{PR} senses the channel through $M$ receive antennas, without any probing signal for transmission.
Fig. \ref{fig_1} depicts a \ac{BS} broadcasting the same signal vector to communication users and an intended target of interest. Communication users are considered to be located at random positions.
In what follows, we present the signal models for the communication and sensing systems.
We note that perfect \ac{CSI} is assumed for all channels.

\subsection{Communication Model}
\label{subsec:comm-model}
Denote $\boldsymbol{S}=\left[\boldsymbol{s}_1, \boldsymbol{s}_2, \cdots, \boldsymbol{s}_{M_t}\right]^{\top} \in \mathbb{C}^{M_t \times L}$ by the transmit waveform matrix of the communication system transmitted by the \ac{BS}, where $\mathbf{s}_m$ represents the transmit signal of the $m^{th}$ antenna element of the \ac{BS}, $ \forall m=1,2, \cdots, M_t$.
In addition, $L$ denotes the number of time samples transmitted by the \ac{BS}.
To this end, the received communication signals can be written as
\begin{equation}
	\label{eq:comm-matrix}
	\boldsymbol{Y}_c=\left(\gamma^{\rm{c}}_{\rm{d}}\boldsymbol{H}_{\rm{k}}+\gamma^{\rm{c}}_{\rm{r}}\boldsymbol{H}_{\rm{R k}} \boldsymbol{\Phi} \boldsymbol{H}_{\rm{c R}}\right) 
	\boldsymbol{S}+
	\boldsymbol{H}_I
	\boldsymbol{S}_I
	+
	\boldsymbol{N}_c,
\end{equation}
where $\boldsymbol{H}_{\rm{k}} \in \mathbb{C}^{M_r \times M_t}$ and $\boldsymbol{H}_{R k} \in \mathbb{C}^{M_r \times N}$ are the normalized channel responses (the small-scale channel matrix) from the \ac{BS} to the user, and from the \ac{RIS} to the user, respectively.
Moreover, $\boldsymbol{H}_{\rm{c R}} \in \mathbb{C}^{N \times M_t}$ represents the normalized channel response from the \ac{BS} to the \ac{RIS}.
In addition, $\boldsymbol{\Phi}=\operatorname{diag}(\boldsymbol{\phi})$ is the reflection matrix of the \ac{RIS}, whereby $\boldsymbol{\phi}=\left[\phi_1, \cdots, \phi_N\right]^{\top}$ is the vector of reflecting coefficients, and $\left|\phi_n\right|=1, \forall n=1, \cdots, N$.
In addition, the matrix $\boldsymbol{S}_I \in \mathbb{C}^{M_r \times L}$ is an interference matrix arising  from various sources, such as co-channel interference. Also $\boldsymbol{H}_I$ is the interferer's channel.
The direct and reflected complex channel gains are denoted as $\gamma^{\rm{c}}_{\rm{d}} \in \mathbb{C}$ and $\gamma^{\rm{c}}_{\rm{r}} \in \mathbb{C}$, respectively, which follow the single and double pathloss equations \cite{goldsmith2005wireless}, \cite{10417003} namely
\begin{equation}
\label{eq:PL-comm}
\begin{split}
\left|\gamma^{\rm{c}}_{\rm{d}}\right|  =\sqrt{\frac{\lambda^2 P_TG_TG_R^{\rm{c}}}{(4 \pi)^2d_{\rm{k}}^2}}, \quad 
\left|\gamma^{\rm{c}}_{\rm{r}}\right|  =\sqrt{\frac{\lambda^2 P_TG_TG_R^{\rm{c}} \sigma_{\mathrm{RIS}}}{(4 \pi)^3d_{\rm{Rk}}^2d_{\rm{cR}}^2}}.
\end{split}
\end{equation}
The transmit power by the \ac{BS} is denoted as $P_T$, 
the \ac{BS}'s transmit antenna gain is $G_T$ and the user's receive antenna gain is $G_R^{\rm{c}}$.
The distances $d_{\rm{k}}$, $d_{\rm{Rk}}$, and $d_{\rm{cR}}$ represent the distance between the \ac{BS} and the user, between the \ac{BS} and the \ac{RIS} and between the \ac{RIS} and user, respectively. 
$\lambda = \frac{c}{f_c}$ is the wavelength, where the carrier frequency is $f_c$ and $c$ is the speed of light.
In the far-field region, the \ac{RCS} of each \ac{RIS} element is as follows \cite{10103813}
\begin{equation}
\label{eq:sigma-ris}
\sigma_{\rm{RIS }}=\frac{4 \pi A^2 S_{\rm{sub }}^2}{\lambda^2} F\left(\varphi_{\mathrm{r}}, \vartheta_{\mathrm{r}}\right) F\left(\varphi_{\mathrm{t}}, \vartheta_{\mathrm{t}}\right),
\end{equation}
where $A$ is the reflecting coefficient of the \ac{RIS} element, which is $1$ in case of passive \ac{RIS}.
Moreover, $S_{\rm {sub}} = d_x \times d_y$ represents the area of each \ac{RIS} element.
$d_x$ and $d_y$ stand for the size of the \ac{RIS} along the $x$ and $y$ dimensions, respectively.
Also, $\left(\varphi^{\mathrm{r}}, \vartheta^{\mathrm{r}}\right)$ and $\left(\varphi^{\mathrm{t}}, \vartheta^{\mathrm{t}}\right)$ denote the azimuth and elevation angles pointing from the center of the \ac{RIS} to the \ac{BS} and target/\ac{PR}. 
In addition, $F\left(\varphi_{\mathrm{r}}, \vartheta_{\mathrm{r}}\right) $ and $F\left(\varphi_{\mathrm{t}}, \vartheta_{\mathrm{t}}\right)$ denote the  normalized power radiation patterns of the \ac{RIS} element in the directions of reflecting and receiving.
The reflecting coefficient of the \ac{RIS} elements is $A = 1$ for passive \ac{RIS}.

Finally, in compact notation, we can write the signal model in equation \eqref{eq:comm-vector} in vectorized form as 
\begin{equation}
	\label{eq:comm-vector}
	\boldsymbol{y}_c=\boldsymbol{H}_c(\phi) \boldsymbol{s}+
	\bar{\boldsymbol{H}}_I
	\boldsymbol{s}_I
	+
	\boldsymbol{n}_c,
\end{equation}
where $\boldsymbol{y}_c=\operatorname{vec}\left(\boldsymbol{Y}_c\right)$, $\boldsymbol{n}_c=\operatorname{vec}\left(\boldsymbol{N}_c\right)$, $\boldsymbol{s}=\operatorname{vec}\left(\boldsymbol{S}\right)$ and $\boldsymbol{s}_I=\operatorname{vec}\left(\boldsymbol{S}_I\right)$ and
\begin{equation}
	\label{eq:Hc_phi}
	\boldsymbol{H}_c(\boldsymbol{\phi})=\boldsymbol{I}_L \otimes \left(\gamma^{\rm{c}}_{\rm{d}}\boldsymbol{H}_{\rm{k}}+\gamma^{\rm{c}}_{\rm{r}}\boldsymbol{H}_{\rm{R k}} \boldsymbol{\Phi} \boldsymbol{H}_{\rm{c R}}\right).
\end{equation}
A similar definition can be made on $\bar{\boldsymbol{H}}_I$.
It is worth noting that the communication channels $\boldsymbol{H}_{\rm{k}},\boldsymbol{H}_{\rm{R k}},\boldsymbol{H}_{\rm{c R}}$ include \ac{LoS} and \ac{NLoS} components that may arise within the wireless propagation. 
In particular, $\boldsymbol{H}_{\rm{k}}$ contains multipath components between the \ac{BS} and the communication user, 
$\boldsymbol{H}_{\rm{R k}}$ may also accommodate any possible static/dynamic obstacles arising between the communication user and the \ac{RIS},
and $\boldsymbol{H}_{\rm{c R}}$ can contain obstacles between the \ac{BS} and the \ac{RIS}.

For the sake of tractability of the analysis below, we shall consider that the interference vector $\boldsymbol{s}_I$ is  Gaussian distributed with variance equal to average interference power, say $\sigma_I^2$. 
This assumption is indeed a \textit{"worst case scenario"}, since, for communications, Gaussian distribution injected onto the signal of interest, yields worst case capacity \cite{8116642,6477131}.
As a result of this assumption, we model $\bar{\boldsymbol{H}}_I\boldsymbol{s}_I$ as an \ac{AWGN} zero-mean and variance $\sigma_I^2$, which represents the average noise power at the user due to interference.
Meanwhile, $\boldsymbol{n}_c$ is also modeled as zero-mean and variance $\sigma_n^2$.
This means that the quantity $\bar{\boldsymbol{H}}_I\boldsymbol{s}_I+\boldsymbol{n}_c$ is itself zero-mean \ac{AWGN} with variance $\sigma_c^2=\sigma_I^2+\sigma_n^2$, i.e. the worst case interference can be viewed as added \ac{AWGN}.

\subsection{Sensing Model}
Referring to Fig. \ref{fig_1}, we can see that the signal at the \ac{PR} arrives from multiple sources.
The first components we model are referred to as \ac{PI}, which include two main components, comprising two primary elements: the \ac{DPI} and the \ac{RPI}. 
As the name suggests, the \ac{DPI} arises due to the direct channel between the \ac{BS} and \ac{PR}.
Denoting $\boldsymbol{H}_{\rm{DPI}} \in \mathbb{C}^{M \times M_t}$ as the normalized \ac{DPI} channel response and $\boldsymbol{H}_{\rm{RPI}} \in \mathbb{C}^{M \times M_t}$ as the normalized \ac{RPI} channel response, we can formally define the \ac{PI} channel as 
\begin{equation}
\label{eq:decompose-PI}
\boldsymbol{H}_{\rm{PI}}
=
\gamma_{\rm{DPI}}
\boldsymbol{H}_{\rm{DPI}}
+
\gamma_{\rm{RPI}}
\boldsymbol{H}_{\rm{RPI}},
\end{equation}
where 
\begin{equation}
\label{eq:RPI}
	\boldsymbol{H}_{\rm{RPI}}
	=
	\boldsymbol{G}_{\rm{r R}}^{H} \boldsymbol{\Phi} \boldsymbol{H}_{\rm{c R}},
\end{equation}
and $\boldsymbol{G}_{\rm{r R}} \in \mathbb{C}^{N \times M}$ represents the normalized channel response from the \ac{PR} to the \ac{RIS}.
Similar to \eqref{eq:PL-comm}, $\gamma_{\rm{DPI}} \in \mathbb{C}$ and $\gamma_{\rm{RPI}} \in \mathbb{C}$ denote the complex channel gains of the \ac{DPI} and \ac{RPI}, respectively, and are given as follows
\begin{equation*}
\begin{split}
\left|\gamma_{\rm{DPI}} \right|  =\sqrt{\frac{\lambda^2 P_TG_TG_R^{\rm{PR}}}{(4 \pi)^2d_{\rm{DPI}}^2}}, \quad 
\left|\gamma_{\rm{RPI}}\right| =\sqrt{\frac{\lambda^2 P_TG_TG_R^{\rm{PR}} \sigma_{\mathrm{RIS}}}{(4 \pi)^3d_{\rm{rR}}^2d_{\rm{cR}}^2}},
\end{split}
\end{equation*}
where all quantities are already defined except for $G_R^{\rm{PR}}$ which is the antenna gain at the \ac{PR} and $d_{\rm{rR}}$ which is the distance between the \ac{RIS} and \ac{PR}.
For a detailed motivation on the importance of mitigating the \ac{PI}, especially in practical scenarios with limited dynamic range, the reader is referred to Section \ref{sec:motivation-of-dynamic-range}.

Furthermore, Fig. \ref{fig_1} illustrates that the useful sensing signal to sense the target from \ac{PR}'s perspective can be modeled through four paths:
$(i)$ In the first path, i.e. \ac{BS} $\rightarrow$ target $\rightarrow$ \ac{PR}, the transmit waveforms hit the target and then echoes back to the \ac{PR}. This channel is denoted as 
	$\gamma_1^{\rm{s}}\boldsymbol{g}_{\rm{t}}\boldsymbol{h}_{\rm{t}}^{H}$, whereby $\boldsymbol{g}_{\rm{t}} \in \mathbb{C}^{M \times 1}$ denotes the target response associated with the direct channel between the target and \ac{PR} and $\boldsymbol{h}_{\rm{t}} \in \mathbb{C}^{M_t \times 1}$ models the channel between the \ac{BS} and the target.
	Also, $\gamma_1^{\rm{s}} \in \mathbb{C}$ is the complex channel gain due to the first path, and follows a double pathloss equation similar to $\gamma^{\rm{c}}_{\rm{r}}$ in equation \eqref{alg:algo1}, but with the exception of $\sigma_t$ to account for the bistatic \ac{RCS} of the target.
$(ii)$ In the second path, which is \ac{BS} $\rightarrow$ target $\rightarrow$ \ac{RIS} $\rightarrow$ \ac{PR}, the transmit signal reaches the target and then backscatters to the \ac{PR} via \ac{RIS}.
This channel can be written as 
	$\gamma_2^{\rm{s}}\boldsymbol{G}_{\rm{r R}}^{H} \boldsymbol{\Phi} \boldsymbol{g}_{\rm{R t}}\boldsymbol{h}_{\rm{t}}^{H}$, where $\boldsymbol{g}_{\rm{R t}} \in \mathbb{C}^{N \times 1}$ represents the normalized channel between the target and the \ac{RIS}.
Furthermore, $\gamma_2^{\rm{s}} \in \mathbb{C}$ is the complex channel gain due to the second path and follows a triple pathloss equation.
$(iii)$ The third path, namely \ac{BS} $\rightarrow$ \ac{RIS} $\rightarrow$ target $\rightarrow$ \ac{PR}, captures the transmit waveforms which hits the \ac{RIS} first. This channel can be expressed as $	\gamma_3^{\rm{s}}\boldsymbol{g}_{\rm{t}}
	\boldsymbol{g}_{\rm{R t}}^{H} 
	\boldsymbol{\Phi} 
	\boldsymbol{H}_{\rm{c R}}$. 
In addition, $\gamma_3^{\rm{s}} \in \mathbb{C}$ is the complex channel gain due to the third path also is in accordance of a triple pathloss equation.
	Finally,
$(iv)$ the fourth path, which is \ac{BS} $\rightarrow$ \ac{RIS} $\rightarrow$ target $\rightarrow$ \ac{RIS} $\rightarrow$ \ac{PR}, the transmit waveforms propagates first towards the \ac{PR}, similar to the third path, but then it is reflected toward the target, and backscattered to the radar receivers. This channel can be expressed as 
	$\gamma_4^{\rm{s}}(\boldsymbol{G}_{\rm{r R}}^{H} \boldsymbol{\Phi} \boldsymbol{g}_{\rm{R t}})
	(
	\boldsymbol{g}_{\rm{R t}}^{H} 
	\boldsymbol{\Phi} 
	\boldsymbol{H}_{\rm{c R}})$.
Moreover, $\gamma_4^{\rm{s}} \in \mathbb{C}$ is the complex channel gain arising due to the fourth path and abides by a quadruple pathloss equation.
For a detailed explanation on the \textit{four-path model}, the reader is referred to \cite{10243495}, \cite{9732186}.
To this end, the signal received at the \ac{PR} can be written as
\begin{equation}
\label{eq:PR-vector}
\begin{split}
	\boldsymbol{Y}_t
	&=
	\boldsymbol{H}_{\rm{PI}} \boldsymbol{S} 
	+
	\big(
	\gamma_1^{\rm{s}}\boldsymbol{g}_{\rm{t}} \boldsymbol{h}_{\rm{t}}^{H}  +
\gamma_2^{\rm{s}}\boldsymbol{G}_{\rm{r R}}^{H} \boldsymbol{\Phi} \boldsymbol{g}_{\rm{R t}} \boldsymbol{h}_{\rm{t}}^{H} +
\gamma_3^{\rm{s}}\boldsymbol{g}_{\rm{t}} \boldsymbol{g}_{\rm{R t}}^{H} \boldsymbol{\Phi} \boldsymbol{H}_{\rm{c R}}\\ &  +
\gamma_4^{\rm{s}}\boldsymbol{G}_{\rm{r R}}^{H} \boldsymbol{\Phi} \boldsymbol{g}_{\rm{R t}} \boldsymbol{g}_{\rm{R t}}^{H} \boldsymbol{\Phi} \boldsymbol{H}_{\rm{c R}}
\big) 
		\boldsymbol{S}
	+
\sum_{i=1}^Q \gamma^{\rm{ob}}_i
	\boldsymbol{g}_{{\rm{ob}},i} \boldsymbol{h}_{{\rm{ob}},i}^{H}
	\boldsymbol{S}
	+
	\boldsymbol{N}_r,
\end{split}
\end{equation} 
where the first term arises due to the \acp{PI}. 
Moreover, the second term factors the four sensing paths.
The third term, i.e. $
\sum_{i=1}^Q \gamma^{\rm{ob}}_i
	\boldsymbol{g}_{{\rm{ob}},i} \boldsymbol{h}_{{\rm{ob}},i}^{H}
	\boldsymbol{S}$ defines the possible ever-changing obstacles found in the scene, e.g. moving clutter \cite{8370234}. More specifically, $\gamma^{\rm{ob}}_i$ is the  complex channel gain arising from the $i^{th}$ obstacle, including its \ac{RCS} and possible doppler shift in the dynamic case. Moreover,
	$\boldsymbol{g}_{{\rm{ob}},i}  \in \mathbb{C}^{M \times 1}$ is the response associated with the direct channel between the $i^{th}$ obstacle and \ac{PR} and $\boldsymbol{h}_{{\rm{ob}},i}\in \mathbb{C}^{M_t \times 1}$ models the channel between the $i^{th}$ obstacle and the \ac{BS}. In addition, $Q$ represents the number of obstacles in the scene.
The last term, i.e. $\boldsymbol{N}_r$ represents the \ac{PR} background noise.\\
In a very compact way, and similar to \eqref{eq:comm-vector}, we vectorize the received signal at the \ac{PR} in \eqref{eq:PR-vector} as
\begin{equation}
\label{eq:PR-vectorized}
	\boldsymbol{y}_t=\boldsymbol{A}_c(\phi) \boldsymbol{s}+\boldsymbol{A}_r(\phi) \boldsymbol{s}+
	\boldsymbol{A}_o(\phi) \boldsymbol{s}+
	\boldsymbol{n}_r,
\end{equation}
where $\boldsymbol{y}_t=\operatorname{vec}\left(\boldsymbol{Y}_t\right)$, $\boldsymbol{n}_r=\operatorname{vec}\left(\boldsymbol{N}_r\right)$ and
\begin{align}
	\label{eq:Ac}
	\boldsymbol{A}_c(\phi)&=\boldsymbol{I}_L \otimes \boldsymbol{H}_{\rm{PI}},
\end{align}
and
{
\begin{equation}
	\label{eq:Ar}
\begin{split}
	\boldsymbol{A}_r(\phi)&=\boldsymbol{I}_L \otimes (
	\gamma_1^{\rm{s}}\boldsymbol{g}_{\rm{t}} \boldsymbol{h}_{\rm{t}}^{H}  +
\gamma_2^{\rm{s}}\boldsymbol{G}_{\rm{r R}}^{H} \boldsymbol{\Phi} \boldsymbol{g}_{\rm{R t}} \boldsymbol{h}_{\rm{t}}^{H}  \\ &+
\gamma_3^{\rm{s}}\boldsymbol{g}_{\rm{t}} \boldsymbol{g}_{\rm{R t}}^{H} \boldsymbol{\Phi} \boldsymbol{H}_{\rm{c R}}  +
\gamma_4^{\rm{s}}\boldsymbol{G}_{\rm{r R}}^{H} \boldsymbol{\Phi} \boldsymbol{g}_{\rm{R t}} \boldsymbol{g}_{\rm{R t}}^{H} \boldsymbol{\Phi} \boldsymbol{H}_{\rm{c R}}
). 
\end{split}
\end{equation}
}
Furthermore, we have that 
\begin{equation}
		\label{eq:Ao}
	\boldsymbol{A}_o(\phi)=\boldsymbol{I}_L \otimes \sum_{i=1}^Q \gamma^{\rm{ob}}_i
	\boldsymbol{g}_{{\rm{ob}},i} \boldsymbol{h}_{{\rm{ob}},i}^{H}.
\end{equation}

In the following section, we highlight the importance of mitigating components arising from \acp{PI} for \ac{ISAC} applications.

\subsection{\ac{ISAC} Central Processing}
\label{subsec:cpu}
A \ac{CPU} could be incorporated to perform the processing, as depicted in Fig. \ref{fig_1}.
At an initial phase, the \ac{CPU} can generate the statistical transmit covariance matrix towards the \ac{BS}, the \ac{RIS} phase shifts, and the space-time analog beamformer to the \ac{PR} through backhauling capacity links, after solving the optimization problems discussed in the coming sections. 
These quantities can then be communicated back to the \ac{BS}, \ac{RIS}, and \ac{PR}, allowing them to make the necessary adjustments.
In the subsequent phase, the \ac{BS} would transmit its data symbols via the obtained covariance matrix, whereas the \ac{PR} performs space-time beamforming for the target detection process.
If a target exists, then the received signal is forwarded back to the \ac{CPU} for further sensing analysis, e.g. delay-Doppler processing, i.e  a range-Doppler map is formed \cite{10200157}, to estimate the range and speed of the target.

\section{Motivation \& Importance of Dynamic Range}
\label{sec:motivation-of-dynamic-range}
\begin{figure*}[!t]
\centering
\subfloat[{Traditional bistatic sensing}]{\includegraphics[width=3.5in]{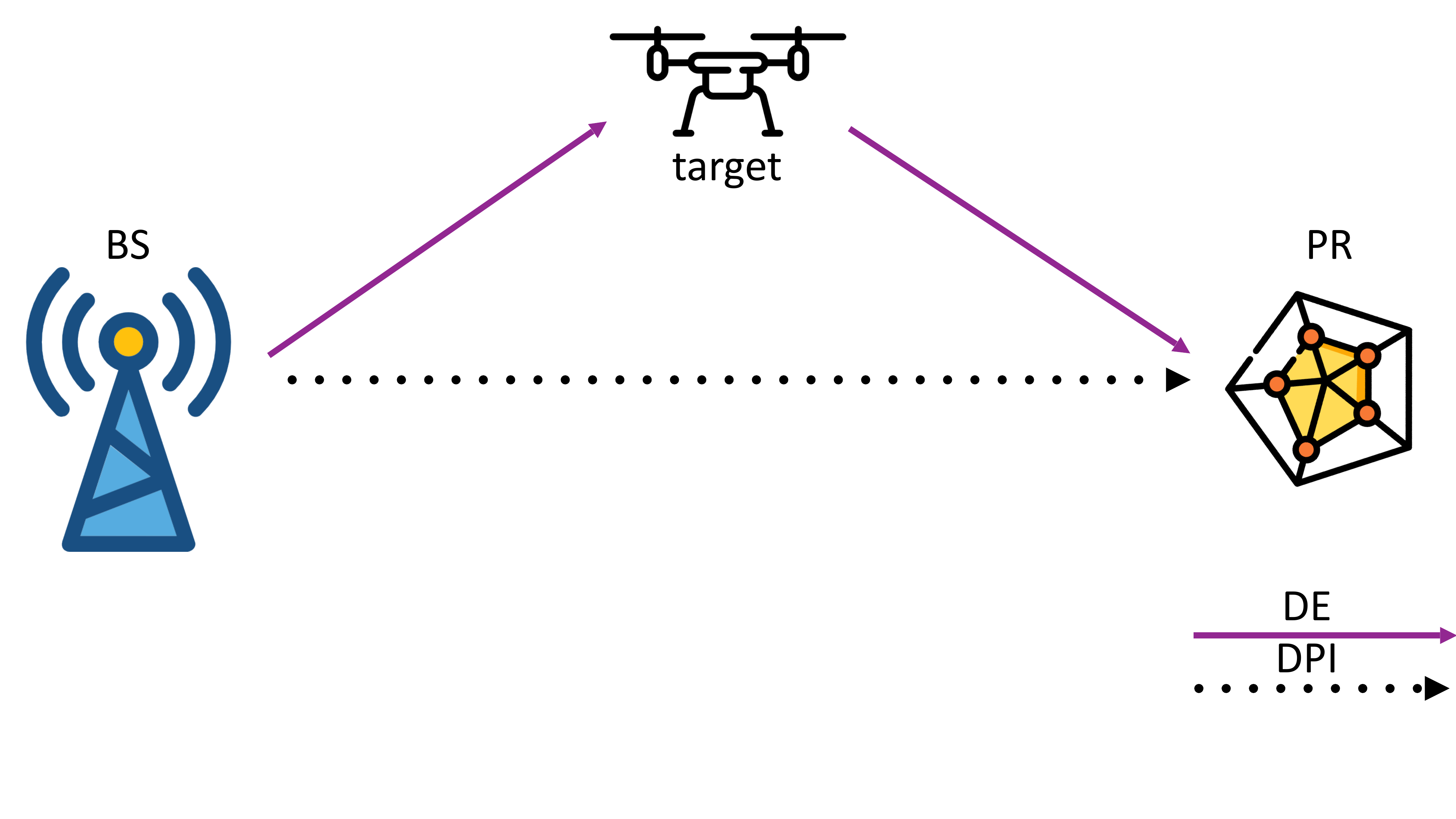} 
\label{fig:fig_DR_bistatic_a}}
\hfil
\subfloat[{\ac{RIS}-aided bistatic sensing}]{\includegraphics[width=3.5in]{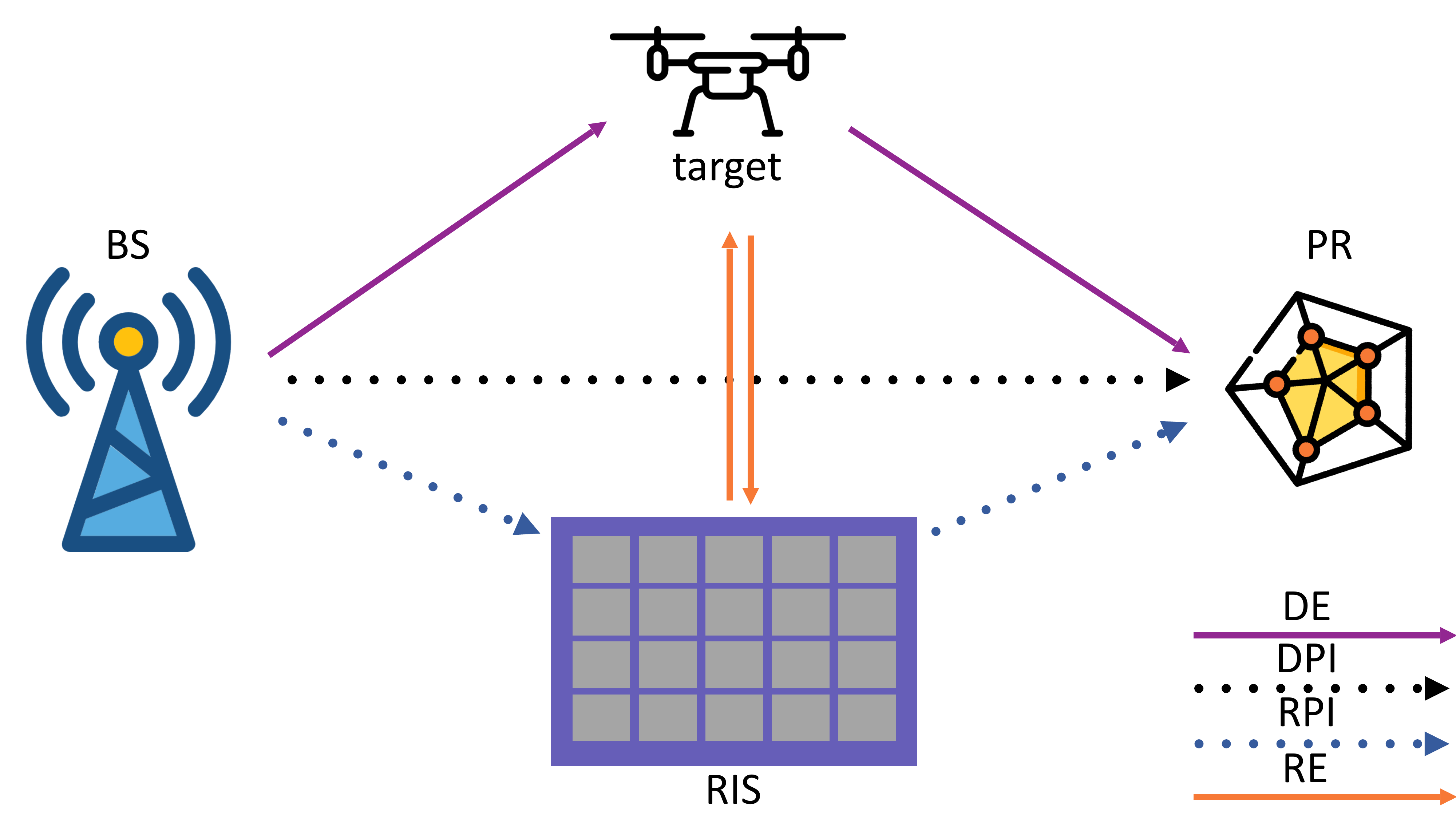}  
\label{fig:fig_DR_bistatic_b}}
\caption{Fundamental difference in \acp{PI} between the traditional and the \ac{RIS}-aided bistatic settings. The setting in (a) contains only a \ac{DPI}, as opposed to \ac{DPI}-plus-\ac{RPI} in (b).}
\label{fig:fig_1}
\end{figure*}
The majority of existing designs neglect the limited dynamic range of practical receivers.
Formally speaking, the \textit{dynamic range} describes the range of the input signal levels that can be reliably measured at the same time.
Said differently, it is the capability to sense small signals in the presence of dominating ones \cite{7397824}.
This issue becomes more pronounced in \ac{ISAC} applications with \ac{RIS} integrations, as strong signals can \textit{"flood"} the \ac{ADC} with un-necessary power.
Indeed, the analog \ac{RF} architecture at the \ac{PR} must be able to ensure proper reception of the weak reflected signals, given the high direct paths total power.
To highlight this, the self-interference caused by the \ac{BS} onto the \ac{PR}, which is referred to in radar literature as \ac{DPI} (also known as \textit{"direct signal interference"} \cite{7891897}), \textit{is most likely multiple orders of magnitude higher than the desired receive signal conveying informative sensing information.}
Putting it in context, the power level gap between these components can exceed $100\dB$, as noted in \cite{griffiths2005passive}.
This, in turn, can overwhelm and saturate components of  the \ac{PR}'s receive chain if not pragmatically mitigated.
In a traditional bi-static scenario, the dynamic range is determined by the ratio of the \ac{DPI} power to the thermal noise level, given that the interest lies in achieving sensitivity down to noise level \cite{malanowski2019signal}.
This arises because, when considering a single target and neglecting clutter, the transmission between transmit and receive units in a bi-static setup can be modeled by only two paths: the \ac{DPI} and the \ac{DE}, as illustrated in Fig. \ref{fig:fig_DR_bistatic_a}.
In contrast, when a \ac{RIS} is present, we can distinguish between four main paths as shown in Fig. \ref{fig:fig_DR_bistatic_b}, two of which are shared with the traditional bi-static scenario (i.e. the \ac{DPI} and \ac{DE}) and two others naturally appearing due to the presence of the \ac{RIS}, namely the \ac{RPI} and the \ac{RE}.
It follows that the dynamic range will not only be determined by the \ac{DPI} power, but also by the \ac{RPI} one, which can also be orders of magnitude stronger than the \ac{DE} and \ac{RE}.
Following \cite{malanowski2019signal}, we can then say that the \ac{ADC}'s dynamic range at the \ac{PR} must satisfy the following condition in order to obtain sensitivity at the level of the thermal noise, i.e. in order to accommodate the \ac{PI} component at a given noise floor 
\begin{equation}
	{\rm DR} \geq  \frac{P_{\rm{PI}}}{P_{\rm{noise}}},
\end{equation}
where $P_{\rm{PI}}$ represents the total power of the \textit{"more general" \ac{PI}} with the useful echo signals, i.e. $P_{\rm{PI}} = P_{\rm{DPI}}+P_{\rm{RPI}}$ where $P_{\rm{DPI}}$ and $P_{\rm{RPI}}$ are the powers of the \ac{DPI} and \ac{RPI} paths, respectively. 
Moreover, $P_{\rm{noise}}$ is thermal noise power which exists in all electronic devices operating at temperatures above absolute zero. 
It is expressed as $P_{\rm{noise}} = k_B T_0 B_r$ \cite{9327501}, \cite{9258405} where 
$k_B$ is the Boltzmann constant,
$T_0$ is the effective noise temperature at the \ac{PR}, and
$B$ is the bandwidth of the \ac{PR}.
A common approach of dealing with \ac{DPI} mitigation (or in this case, the \ac{PI}) is through analog suppression \cite{7891897}, \cite{9764297}, \cite{8513747}, which could cancel certain portions of the \ac{DPI} power.
The main reason of this delegation is because the cancellation performance in the digital domain is limited by the dynamic range of the \ac{ADC} at the \ac{PR}.
Of course, if we were fortunate enough to select an \ac{ADC} with infinite resolution, there would be no need to allocate part of this suppression to the analog domain.
Nevertheless, the dynamic range of currently available \acp{ADC} is insufficient to capture the power difference between the \ac{PI} and all pertinent echoes, resulting in the loss of the desired received signal during the quantization process.
Therefore, jointly reducing the powers of \ac{DPI} and \ac{RPI} can significantly reduce the dynamic range subsequently required by an embedded \ac{ADC} within the receive path of the \ac{PR}.
The \ac{ADC}'s dynamic range is also a measure of the maximum \ac{SNR} \cite{4653945} by the \ac{ADC} \cite{7130672},
which can be computed by the following formula \cite{5672380},\cite{van2013cmos}
\begin{equation}
	{\rm SNR}= 6.02 N_{\rm{ENOB}}+1.76 \ {\rm dBFS},
\end{equation}
where $N_{\rm{ENOB}}$ is the \textit{effective number of bits} in the ADC resolution.
For practicality reasons, it should be noted that the effective number of bits is the number of bits required by non-ideal \acp{ADC}, such as the sigma-delta \ac{ADC} \cite{482138}, to achieve the same effective resolution as an ideal \ac{ADC} sampling at a resolution of $N_{\rm{ENOB}}$ \cite{5672380}.
In fact, the effective number of bits is upper bounded by the actual number of bits of an \ac{ADC}, as the former contains further non-idealities \cite{morgan2012comparison}, such as clipping, jitter and non-linearities.
One example is the $14$-bit AD $9683$ \cite{ad968314} targeting communication applications, but an effective number of bits of around $11$bits.
Keep in mind that the quantizer is the main contributor to power consumption in an \ac{ADC}. Consequently, having more bits per measurement directly leads to slower sampling rates and more expensive \ac{ADC} costs \cite{6418031}. 
Technically speaking, the dissipated \ac{ADC} power is given as $P_{\mathrm{ADC}}= \frac{2^{b} f_{\rm{samp}}}{F} $, where $b$ is the number of bits used by the \ac{ADC}'s quantizer,
$f_{\rm{samp}}$ is the \ac{ADC}'s sampling rate, and
$F$ is the figure of merit \cite{761034}.
Hence, increasing $b$ exponentially increases the required \ac{ADC} power \cite{7227037}, \cite{7953405}.
Therefore, suppression of the \ac{PI} prior to any \ac{PR} signal processing or detection scheme is crucial for optimizing the effective number of bits, 
in order to improve radar detection capabilities and enhance the overall performance of the system.
%
%
%
%
%
%
%
%
%
%
%
%

\section{\ac{RIS}-\ac{ISAC} Optimization with Reduced Dynamic Range}
\label{sec:dynamic-range-minimization}
\subsection{Problem Statement}
Assuming the \ac{PR} is equipped with a space-time analog beamformer, i.e. $\pmb{w}
=
\begin{bmatrix}
	w_1 & \ldots & w_{LM}
\end{bmatrix}^T
 \in \mathbb{C}^{LM \times 1}$. 
More precisely, the space-time analog beamformer $\pmb{w}$ is applied before \ac{ADC} conversion.
Using the system model in equation \eqref{eq:PR-vectorized}, we can decompose the powers as follows
\begin{equation}
\label{eq:aBFing}
P_{\rm{aBFing}}
=
\mathbb{E}( \vert {\pmb{w}}^H \pmb{y}_t \vert^2)
=
P_{\rm{PI}}
+
P_{\rm{sense}}
+
P_{\rm{obs}}
+
P_{\rm{noise}},
\end{equation}
where $P_{\rm{PI}}$ is the power of the \ac{PI} over the observation time,
$P_{\rm{sense}}$ is the power of the signal part containing useful sensing information at the \ac{PR}, and 
$P_{\rm{obs}}$ represents the power due to reflections from obstacles over the observation time.Also, $P_{\rm{noise}}$ is the power of the noise at the \ac{PR}. The power quantities can be verified to be 
\begin{align}
	\label{eq:P_PI}
	P_{\rm{PI}} &= {\pmb{w}}^H\boldsymbol{A}_c(\phi) \pmb{R}_{ss} 
	\boldsymbol{A}_c^H (\phi) 
	{\pmb{w}}, \\
	\label{eq:P_sense}
	P_{\rm{sense}} &= {\pmb{w}}^H\boldsymbol{A}_r(\phi) \pmb{R}_{ss} 
	\boldsymbol{A}_r^H (\phi) 
	{\pmb{w}}, \\
	\label{eq:P_obs}
	P_{\rm{obs}} &= {\pmb{w}}^H\boldsymbol{A}_o(\phi) \pmb{R}_{ss} 
	\boldsymbol{A}_o^H (\phi) 
	{\pmb{w}},\\
	\label{eq:P_noise}
	P_{\rm{noise}} &= \sigma_r^2 \Vert {\pmb{w}} \Vert^2. 
\end{align}
where $\boldsymbol{R}_{ss}=\mathbb{E}\left(\boldsymbol{s s}^{H}\right)$ is the statistical covariance matrix of the transmit symbols $\pmb{s}$. 
At this point, the \ac{PR}'s \ac{KPI} can be defined through the \ac{SNDR} (which can also be interpreted as an \ac{SINR} \cite{7891897}), which is given by \cite{7891897}(c.f. eqn. (10))
\begin{equation}
\label{eq:SNDR_equation}
	{\rm SNDR} = \frac{P_{\rm{sense}}}{P_{\rm{PI}} +P_{\rm{obs}} + P_{\rm{noise}}}.
\end{equation}
Note that the \ac{PI} has been accommodated as part of interference at the \ac{PR}, as some \ac{PI} can still be found in the signal model even after suppression (namely the residual component).
A remark here is that, compared to \cite{7891897}(c.f. eqn. (10)), we have accounted for possible obstacles that can arise as interferers within the sensing \ac{SNDR}.
Now that all quantities are defined, we can clearly formulate our collaborative \ac{BS}, \ac{RIS} and \ac{PR} optimization problem as follows
\begin{equation}
\label{eq:main-problem}
	(\mathcal{P})\left\{\begin{aligned}
\min _{\boldsymbol{w}, \boldsymbol{R}_{ss}, \boldsymbol{\phi}} \quad  & P_{\rm{PI}} \\
\rm { s.t. } \quad  
& \frac{\operatorname{tr}\left(\boldsymbol{R}_{ss} \boldsymbol{H}_c^{H}(\boldsymbol{\phi}) \boldsymbol{H}_c(\boldsymbol{\phi})\right)}{M_r L \sigma_c^2} \geq \gamma_{\rm{comm}}, \\
& \frac{P_{\rm{sense}}}{P_{\rm{PI}}+P_{\rm{obs}}+P_{\rm{noise}}} \geq \gamma_{\rm{sense}},\\
& |\phi_n|=1, \quad n=1 \cdots N, \\
& |w_k|=1, \quad  k=1 \cdots LM, \\
& \operatorname{tr}\left(\boldsymbol{R}_{ss}\right)=P_B, \quad \boldsymbol{R}_{ss} \succeq \pmb{0}.
\end{aligned}\right.
\end{equation}
The proposed problem $(\mathcal{P})$ aims at minimizing the total \ac{PI} power with only analog space-time processing at the level of the \ac{PR} and \ac{RIS} (namely through $\pmb{w}$ and $\pmb{\phi}$ configuration) and digital processing at the level of the \ac{BS}, i.e. via $\pmb{R}_{ss}$ optimization.
Moreover, the first constraint is the communication \ac{SNR}, whereby $\gamma_{\rm{comm}}$ is the desired communication \ac{SNR} to satisfy, which is given as input to the problem.
Likewise, the second constraint guarantees a certain radar \ac{SNDR}, where $\gamma_{\rm{sense}}$ is the minimum accepted target radar \ac{SNDR} to guarantee.
Note that the radar \ac{SNR} (in this case radar \ac{SNDR}) is a widely spread metric \cite{8917703}, \cite{8678798}.
The third and fourth constraints ensure constant-modulus solutions onto the \ac{RIS} phase shifts and the analog space-time beamformer, respectively. 
The last constraint imposes a power constraint, where $P_B$ is the available power budget. In addition, the last constraint imposes a positive definite constraint on the statistical covariance matrix of the transmit symbols.
The problem $(\mathcal{P})$ is non-convex and difficult to solve in its current form.
The non-convexity arises due to the constant-modulus constraints \cite{9857946}.
In the following subsection, we devise a \ac{BCCD} tailored specifically for problem $(\mathcal{P})$ in \eqref{eq:main-problem}.

\subsection{Optimization via Block Cyclic Coordinate Descent Design}
\label{subsec:opt-via-block-cyclic}
First, we solve a subproblem of equation \eqref{eq:main-problem}, which is 
\begin{equation}
\label{eq:main-subproblem}
	(\mathcal{P})\left\{\begin{aligned}
\min _{\boldsymbol{w}, \boldsymbol{\phi}} & \  P_{\rm{PI}} \\
\rm { s.t. } \quad  
& |\phi_n|=1, \quad n=1 \cdots N, \\
& |w_k|=1, \quad  k=1 \cdots LM.
\end{aligned}\right.
\end{equation}
Note that the above requires knowledge, or at least an estimate, of $\pmb{R}_{ss}$.
In the first iteration, a random initializer of $\pmb{R}_{ss}$ suffices.
Before we proceed, we re-write the objective function $P_{\rm{PI}}$, which gives us the following equivalent optimization problem 
\begin{equation}
\label{eq:sub-problem-1}
	(\mathcal{P}_1) \left\{\begin{aligned}
\min _{\boldsymbol{w}, \boldsymbol{\phi}} & \  \sum_{i=1}^{L M_t} \vert \pmb{w}^H \pmb{b}_i +  {\pmb{w}}^H \pmb{C}_i\boldsymbol{\phi} \vert^2 \\
\rm { s.t. } \quad  
& |\phi_n|=1, \quad n=1 \cdots N, \\
& |w_k|=1, \quad  k=1 \cdots LM.
\end{aligned}\right.
\end{equation}
where 
{
\begin{align}
	\label{eq:bi}
	\pmb{b}_i &= 
	 \sqrt{\lambda}_i \gamma_{\rm{DPI}} (\boldsymbol{I}_L \otimes \boldsymbol{H}_{\rm{DPI}}) \pmb{v}_i \\
	 \label{eq:Ci}
	 \pmb{C}_i
	&=
	\sqrt{\lambda}_i 
	 \gamma_{\rm{RPI}}
	 \boldsymbol{G}_{\rm{r R}}^{H}
	\begin{bmatrix}
		 \operatorname{diag}(\boldsymbol{H}_{\rm{c R}}\pmb{v}_i^{(1)})  \hdots  
		 \operatorname{diag}(\boldsymbol{H}_{\rm{c R}}\pmb{v}_i^{(L)})
	\end{bmatrix},
\end{align}
}
where $\pmb{v}_i = \begin{bmatrix}
	(\pmb{v}_i^{(1)})^T& \ldots & (\pmb{v}_i^{(L)})^T
\end{bmatrix}^T$ and $\pmb{v}_i^{(\ell)} \in \mathbb{C}^{M_t \times 1}$ for $\ell = 1 \ldots L$.
The details are found in \textbf{Appendix \ref{app:problem-equivalence}}.
As the problem in $(\mathcal{P}_1)$ is still non-convex, it is easier to tackle than the original problem in $(\mathcal{P})$, which can be solved through an \ac{RCG} algorithm.  
Before we proceed, we unify the variables into one variable
$
	\pmb{x} 
	=
	\begin{bmatrix}
		\pmb{w}^T &  \pmb{\phi}^T
	\end{bmatrix}^T = \begin{bmatrix}
		x_1 & \ldots & x_{N+LM}
	\end{bmatrix}^T
$, hence we can easily re-write the problem as 
\begin{equation}
 \label{eq:CancelPR-RIS-x}
	(\mathcal{P}_1) \left\{\begin{aligned}
\min _{\boldsymbol{w}, \boldsymbol{\phi}} & \  \sum_{i=1}^{L M_t} \vert \pmb{x}^H \bar{\pmb{b}}_i +  {\pmb{x}}^H \pmb{D}_i\pmb{x} \vert^2 \\
\rm { s.t. } \quad  
& |{x}_j|=1, \quad j=1 \cdots N+LM, 
\end{aligned}\right.
\end{equation}
where $\bar{\pmb{b}}_i$ is a zero-padded version of ${\pmb{b}}_i$ as follows $\bar{\pmb{b}}_i = 
\begin{bmatrix}
	\bar{\pmb{b}}_i^T & 
	\pmb{0}_{1 \times N}^T
\end{bmatrix}^T$ and 
\begin{equation}
	\pmb{D}_i
	= 
	\begin{bmatrix}
		\pmb{I}_{ML} \\
		\pmb{0}_{N \times ML}
	\end{bmatrix}
	\pmb{C}_i
	\begin{bmatrix}
		\pmb{0}_{N \times ML} & \pmb{I}_{N} 
	\end{bmatrix}.
\end{equation}
The Riemannian gradient, $\grad f(\pmb{x}_i)$, at any given point $\pmb{x}_i$ is defined as the orthogonal projection of the Euclidean gradient $\nabla f(\pmb{x}_i)$ onto the tangent space $T_{\pmb{x}_i} \mathcal{M}$ of the manifold $\mathcal{M}$ at point $\pmb{x}_i$. This projection can be mathematically expressed
\begin{equation}
	T_{\pmb{x}_i} \mathcal{M}
	=
	\lbrace
	\pmb{y} \in \mathbb{C}^{M +  N_{\tt{PR}}}
	\vert 
	\Real(
	\pmb{y}
	\odot
	\pmb{x}_i^*
	)
	=
	\pmb{0}_{N_{\tt{PR}}}
	\rbrace.
\end{equation}
Considering the objective function in hand, i.e. $f(\pmb{x}) = \sum_{i=1}^{L M_t} \vert \pmb{x}^H \bar{\pmb{b}}_i +  {\pmb{x}}^H \pmb{D}_i\pmb{x} \vert^2$, the gradient in the Euclidean space at the point $\pmb{x}$ is expressed as
\begin{equation}
	\label{eq:nabla_f_x}
	\nabla f(\pmb{x})
	=
	2
	\sum_{i=1}^{L M_t}
	\begin{bmatrix}
	(\pmb{C}_i\pmb{\phi} + \pmb{b}_i)(\pmb{\phi}^H\pmb{C}_i^H\pmb{w} + (\pmb{w}^H\pmb{b}_i)^*) \\
	\pmb{C}_i^H \pmb{w}(\pmb{w}^H \pmb{C}_i \pmb{\phi}  + \pmb{w}^H \pmb{b}_i)
	\end{bmatrix},
\end{equation}
where we have expressed the gradient in terms of $\pmb{w}$ and $\pmb{\phi}$.
Furthermore, the Riemannian gradient at $\pmb{x}$ is given as 
\begin{equation}
\label{eq:riem-grad}
	\pmb{g} = 
	\grad f(\pmb{x})
	=
	\nabla f(\pmb{x})
	-
	\Real(
	\nabla f(\pmb{x})
	\odot
	\pmb{x}^*
	)
	\odot
	\pmb{x}.
\end{equation}
Note that utilizing the Riemannian gradient allows extending optimization techniques from Euclidean space to manifold space.
In addition, we would like to mention that the Armijo-Goldstein (or the backtracking line-search) is enforced as a first step of each iteration to ensure the monotonic decrease of $f(\pmb{x})$ as $\pmb{x}_i$ is updated to $\pmb{x}_{i+1}$.
Moreover, the retraction step is defined as 
\begin{equation}
\begin{split}
	\Retr_{\pmb{x}} &: T_{\pmb{x}} \mathcal{M} \rightarrow \mathcal{M} \\ 
	&:\alpha \pmb{c} \rightarrow \Retr_{\pmb{x}}(\alpha \pmb{c}),
\end{split}
\end{equation}
where $\Retr_{\pmb{x}}(\alpha \pmb{c})$ is an $M \times 1$ vector given as
\begin{equation}
\label{eq:re_tr_a_c_m}
	[\Retr_{\pmb{x}}(\alpha \pmb{c})]_m
	=
	\frac{[\pmb{x} + \alpha \pmb{c}]_m}{\vert [\pmb{x} + \alpha \pmb{c}]_m \vert}.
\end{equation}
Since $\pmb{c}_{i+1}$ and $\pmb{c}_i$ fall in different tangent spaces, hence the transport vector is introduced to map the previous search direction from its original tangent space to the current tangent space at the updated point $\pmb{x}_{i+1}$.
For this, we define the transport vector as
\begin{equation}
\label{eq:transport}
\begin{split}
	\mathcal{T}_{\pmb{x}_i \rightarrow \pmb{x}_{i+1}}
	&:
	T_{\pmb{x}_i}\mathcal{M}
	\rightarrow
	T_{\pmb{x}_{i+1}}\mathcal{M} \\
	&: 
	\pmb{c} \rightarrow \pmb{c}
	-
	\Real(
	\pmb{c} \odot \pmb{x}_{i+1}^* 
	)
	\odot \pmb{x}_{i+1}.
\end{split}
\end{equation}
The \ac{RCG} algorithm tailored to solve sub-problem $(\mathcal{P}_1)$ in equation \eqref{eq:sub-problem-1} is summarized in \textbf{Algorithm \ref{alg:algo1}}.
Referring to \textbf{Algorithm \ref{alg:algo1}}, we can observe that the method operates of $I$ iterations, which is an input parameter of the algorithm, and is related to the final obtained precision of the solution.

\begin{algorithm}
\caption{\ac{RCG}-based optimization to solve $(\mathcal{P}_1)$}\label{alg:algo1}
\begin{algorithmic}[1]
\State \textbf{INPUT} $\lbrace \pmb{b}_i,\pmb{C}_i \rbrace_{i=1}^{LM_t}$, $I$
\State Initialize $\pmb{c}_0 = -\grad(\pmb{x}_0)$.
\State Set $i \gets 0 $.
\While{ $i < I$ }
\State Select step size $\alpha_i$ based on backtracking line-search.
\State Update $\pmb{x}_{i+1}$ via Retraction, i.e. $\pmb{x}_{i+1} = \Retr_{\pmb{x}_i}(\alpha_i \pmb{c}_i)$.
\State Compute the Riemannian gradient $\pmb{g}_{i+1} = \grad f(\pmb{x}_{i+1})$ following \eqref{eq:riem-grad}
\State Compute transport vectors $\pmb{g}_i^+ = \mathcal{T}_{\pmb{x}_i \rightarrow \pmb{x}_{i+1}}(\pmb{g}_i)$ in \eqref{eq:transport}.
\State Compute transport vectors $\pmb{c}_i^+ = \mathcal{T}_{\pmb{x}_i \rightarrow \pmb{x}_{i+1}}(\pmb{c}_i)$ in \eqref{eq:transport}.
\State Update $\beta_{i+1} = \frac{\Vert \pmb{g}_{i+1} \Vert^2}{\Vert \pmb{g}_{i}^+ \Vert^2}$
\State Update conjugate gradient as $\pmb{c}_{i+1} = - \pmb{g}_{i+1} + \beta_{i+1} \pmb{c}_{i+1}^+$
\State $i \gets i + 1$
\EndWhile

\State \Return $\pmb{x}_{I} = 	\begin{bmatrix}
		\pmb{w}_I \\ \pmb{\phi}_I
	\end{bmatrix}$ 
\end{algorithmic}
\end{algorithm}

Next, given the current estimates of $\pmb{w}$ and $\pmb{\phi}$, we can formulate a complementary sub-problem to solve for the statistical covariance matrix of the transmit symbols, which is the following
\begin{equation}
\label{eq:main-subproblem-2}
	(\mathcal{P}_2)\left\{\begin{aligned}
\min _{\boldsymbol{R}_{ss}} \quad  & P_{\rm{PI}} \\
\rm { s.t. } \quad  & 
 \frac{\operatorname{tr}\left(\boldsymbol{R}_{ss} \boldsymbol{H}_c^{H}(\boldsymbol{\phi}) \boldsymbol{H}_c(\boldsymbol{\phi})\right)}{M_r L \sigma_c^2} \geq \gamma_{\rm{comm}}, \\
& \frac{P_{\rm{sense}}}{P_{\rm{PI}}+P_{\rm{obs}}+P_{\rm{noise}}} \geq \gamma_{\rm{sense}}, \\
& \operatorname{tr}\left(\boldsymbol{R}_{ss}\right)=P_B, \quad \boldsymbol{R}_{ss} \succeq \pmb{0}. \\
\end{aligned}\right.
\end{equation}
With the help of equations, \eqref{eq:P_PI}, \eqref{eq:P_sense}, \eqref{eq:P_obs}, and \eqref{eq:P_noise}, 
the problem in \eqref{eq:main-subproblem-2} can be equivalently formulated in terms of traces as given at the bottom of the next page in equation \eqref{eq:main-subproblem-3}.
 \begin{figure*}[b]
\input{sections/problemP2.tex}
\end{figure*}
The equivalent problem in equation \eqref{eq:main-subproblem-3} is a convex optimization problem, and in particular an \ac{SDP} problem, which is computationally tractable \cite{9640472} and can be efficiently solved by off-the-shelf optimization software packages, such as MOSEK \cite{andersen2000mosek} or CVX \cite{grant2014cvx}.
The \ac{BCCD} algorithm designed to solve problem $(\mathcal{P})$ in equation \eqref{eq:main-problem} is summarized in \textbf{Algorithm \ref{alg:algo2}}.
Referring to \textbf{Algorithm \ref{alg:algo2}}, $N_{\mathrm{iter}}$ refers to the number of iterations of the main loop to be executed. 
Moreover, we have initialized $\boldsymbol{R}_{s s}^{(0)} = \pmb{U}^H \pmb{U}$, where $\pmb{U}$ is uniformly random matrix of size $LM_t \times LM_t$.
This initialization was chosen to preserve the positive semi-definiteness of the covariance matrix $\boldsymbol{R}_{s s}^{(0)}$.

We would like to highlight that an interferers for communications and obstacles for sensing impact the performance in a crucial way.
For communications, we can see that a higher level of average noise power at the user due to interference, i.e. $\sigma_I^2$ contributes to higher $\sigma_c^2$, thus deteriorating the communication \ac{SNR} in, i.e. $\frac{\operatorname{tr}\left(\boldsymbol{R}_{ss} \boldsymbol{H}_c^{H}(\boldsymbol{\phi}) \boldsymbol{H}_c(\boldsymbol{\phi})\right)}{M_r L \sigma_c^2}$ appearing in the constraints of \eqref{eq:main-problem}. This means that more interference can lead to a less feasible $\gamma_{\mathrm{comm}}$.
For sensing, obstacles lead to a lower \ac{SNDR}, hence impacting the possibility of targeting higher $\gamma_{\mathrm{sense}}$. In fact, the added power from obstacles, $P_{\rm{obs}}$, can be viewed as additional un-necessary power.

\begin{algorithm}
\caption{\ac{BCCD}-based optimization to solve $(\mathcal{P})$}\label{alg:algo2}
\begin{algorithmic}[1]
\State \textbf{INPUT}:
$\gamma_{\rm{comm}}$, $\gamma_{\rm{sense}}$, $\sigma_r^2$, $\sigma_c^2$, $I$, $N_{\mathrm{iter}}$, $P_B$
\State Initialize $\boldsymbol{R}_{ss}^{(0)}$.
\State Set $n \gets 1 $.
\While{ $n < N_{\mathrm{iter}}$ }
\State Perform an \ac{EVD} on $\pmb{R}_{ss}^{(n-1)}$.
\State Given $\pmb{R}_{ss}^{(n-1)}$, compute $\lbrace \pmb{b}_i,\pmb{C}_i \rbrace_{i=1}^{LM_t}$ following \eqref{eq:bi} and \eqref{eq:Ci}.
\State Given $\lbrace \pmb{b}_i,\pmb{C}_i \rbrace_{i=1}^{LM_t}$, run \textbf{Algorithm \ref{alg:algo1}} to estimate $\pmb{w}^{(n)}$ and $\pmb{\phi}^{(n)}$.
\State Given $\pmb{w}^{(n)}$ and $\pmb{\phi}^{(n)}$, solve $(\mathcal{P}_2)$ in \eqref{eq:main-subproblem-3} to obtain $\pmb{R}_{ss}^{(n)}$.
\State $n \gets n + 1$
\EndWhile

\State \Return $\boldsymbol{w}, \boldsymbol{R}_{ss}, \boldsymbol{\phi}$
\end{algorithmic}
\end{algorithm}


\subsection{Computational Complexity}
\label{subsec:complexity-analysis}
\subsubsection{Complexity of the \ac{RCG}-based optimization}
First, we analyze the computational complexity of Algorithm \ref{alg:algo1}. 
Observe that we can decompose the complexity based on the associated updates, namely
\begin{equation}
	\label{eq:T_algo_1}
	T_{\rm{Alg},1}
	=
	I
	\left(
	T_{\alpha}
	+
	T_{\pmb{x}}
	+
	T_{\pmb{g}}
	+
	T_{\pmb{g}^+}
	+
	T_{\pmb{c}^+}
	+
	T_{\beta}
	+
	T_{\pmb{c}}
	\right),
\end{equation}
where 
$T_{\alpha}$ is the computational complexity of selecting step size $\alpha_i$ through backtracking line-search,
$T_{\pmb{x}}$ is the complexity of the retraction step, and
$T_{\pmb{g}},
	T_{\pmb{g}^+},
	T_{\pmb{c}^+},
	T_{\beta},
	T_{\pmb{c}}$ represent the complexities of updating $\pmb{g}_{i+1},\pmb{g}_{i}^+,\pmb{c}_{i}^+, \beta_{i+1},$ and $\pmb{c}_{i+1}$, respectively.
The backtracking line-search requires $\mathcal{O}(\log (1 / \epsilon_1))$ iterations to generate a solution with accuracy $\epsilon_1 > 0$ \cite{boyd2004convex,10106790}.
Moreover, each iteration involves computing the function $f$, which costs $\mathcal{O}\left(LM_t(ML+N)^2\right)$, which stems from the quadratic term, so, {$T_{\alpha}  = \mathcal{O}\left(LM_t(ML+N)^2 \log \frac{1}{\epsilon_1} \right)$}.
The retraction step entails $LM + N$ divisions, hence a total cost of  {$ T_{\pmb{x}}=\mathcal{O}(LM + N)$}.
Updating $\pmb{g}_{i+1}$ necessitates the computation of $\nabla f(\pmb{x})$ first.
The major complexity involved in each summand term in \eqref{eq:nabla_f_x} is due to 
$\pmb{C}_i \pmb{\phi}$, which costs $\mathcal{O}(LMN)$ and $\pmb{w}^H \pmb{b}_i$ which requires $\mathcal{O}(LM)$.
This means that the number of operations involved in computing $\nabla f(\pmb{x})$ is $\mathcal{O}\left( LM_t \left(  LMN + LM \right)\right) =\mathcal{O}\left( L^2 M_t MN\right)$.
After computing the gradient $\nabla f(\pmb{x})$, we notice that \eqref{eq:riem-grad} involves two Hadamard operations of vectors of sizes $LM + N$, hence a complexity of $\mathcal{O}(LM+N)$.
We conclude that  {$T_{\pmb{g}} = \mathcal{O}\left( L^2 M_t MN \right)$} as $L^2 M_t MN \gg LM+N$ is the dominating term.
The number of flops to compute the transport vector $\pmb{g}_i^+$ following \eqref{eq:transport} involves two Hadamard operations followed by a subtraction, which is $3(LM + N)$, therefore the order of complexity translates to  
 {$T_{\pmb{g}^+}=\mathcal{O}(LM+N)$}.
Note that $T_{\pmb{c}^+} = T_{\pmb{g}^+}$, hence  {$T_{\pmb{c}^+} = \mathcal{O}(LM+N)$}.
The update of $\beta_i$ consists of two vector magnitude computations, $\Vert \pmb{g}_{i} \Vert^2$ and $\Vert \pmb{g}_{i+1} \Vert^2$, each of which cost cost $\mathcal{O}(LM+N)$. This is followed by a division operation which is $\mathcal{O}(1)$. We conclude that  {$T_{\beta} = \mathcal{O}(LM+N)$}.
Moreover, computing $\pmb{c}_{i+1}$ demands $LM+N$ additions and $LM+N$ multiplications, so  {$T_{\pmb{c}} = \mathcal{O}(LM+N)$}.
Replacing all obtained complexities of $T_{\alpha},
	T_{\pmb{x}},
	T_{\pmb{g}},
	T_{\pmb{g}^+},
	T_{\pmb{c}^+},
	T_{\beta},
	T_{\pmb{c}}$ in \eqref{eq:T_algo_1} and by neglecting the non-dominating terms abiding by the definition of \textit{"big-oh"},   we conclude that the complexity of \textbf{Algorithm \ref{alg:algo1}} is 
\begin{equation}
	\label{eq:final-complexity-of-alg1}
	T_{\rm{Alg},1}
	=
	\mathcal{O}
	\left(
	LM_t(ML+N)^2 I \log (1 / \epsilon_1)
	+
	L^2 M_t MN I
	\right).
\end{equation}
Next, we conduct a complexity analysis on \textbf{Algorithm \ref{alg:algo2}}.

\subsubsection{Complexity of the BCCD-based optimization}
Following the analysis of \textbf{Algorithm \ref{alg:algo1}}, we now conduct a complexity analysis of \textbf{Algorithm \ref{alg:algo2}}, which can be decomposed as 
\begin{equation}
	\label{eq:T_algo_2}
	T_{\rm{Alg},2}
	=
	N_{\rm{iter}}
	\left(
	T_{\rm{EVD}}
	+
	T_{\pmb{b}}
	+
	T_{\pmb{C}}
	+
	T_{\rm{Alg},1}
	+
	T_{\mathcal{P}_2}
	\right),
\end{equation}
where $T_{\rm{EVD}}$ is the cost to perform the \ac{EVD} on $\pmb{R}_{ss}^{(n-1)}$, $T_{\pmb{b}},
	T_{\pmb{C}}$ are the costs of computing all $\lbrace \pmb{b}_i\rbrace_{i=1}^{LM_t}$ and $\lbrace \pmb{C}_i \rbrace_{i=1}^{LM_t}$, respectively. 
Moreover, $T_{\mathcal{P}_2}$ is the cost to solve $(\mathcal{P}_2)$.
The complexity in decomposing $\pmb{R}_{ss}^{(n-1)}$ via \ac{EVD} is  {$T_{\rm{EVD}} = \mathcal{O}(L^3M_t^3)$}.
Each $\pmb{b}_i$ in \eqref{eq:bi} consists of a matrix multiplication of $\boldsymbol{I}_L \otimes \boldsymbol{H}_{\rm{DPI}}$ with  $\pmb{v}_i$, 
which without any structure exploitation would yield $\mathcal{O}(L^2M M_t)$, however, it should be pointed out that  utilizing a peculiar structure of the Kronecker product in this case allows us to write 
\begin{equation}
\label{eq:special-struct-of-kron}
	\left( \boldsymbol{I}_L \otimes \boldsymbol{H}_{\rm{DPI}} \right) \pmb{v}_i 
	=
	\begin{bmatrix}
		\boldsymbol{H}_{\rm{DPI}}\pmb{v}_i^{(1)}  \\ 
		 \vdots \\  \boldsymbol{H}_{\rm{DPI}}\pmb{v}_i^{(L)} 
	\end{bmatrix},
\end{equation} 
where $\pmb{v}_i^{(\ell)} 
$ is defined below \eqref{eq:Ci}.
Each operation of type $\boldsymbol{H}_{\rm{DPI}}\pmb{v}_i^{(\ell)} $ costs $\mathcal{O}(MM_t)$, hence by \eqref{eq:special-struct-of-kron}, the multiplication costs $\mathcal{O}(LMM_t)$ and we conclude that  {$T_{\pmb{b}} = \mathcal{O}(L^2MM_t^2)$}.
Each row of the matrix appearing in \eqref{eq:Ci} consists of a multiplication $\boldsymbol{H}_{\rm{c R}}\pmb{v}_i^{(\ell)}$, which costs $\mathcal{O}(NM_t)$, $\forall \ell$. Moreover, $\boldsymbol{G}_{\rm{r R}}^{H} \operatorname{diag}(\boldsymbol{H}_{\rm{c R}}\pmb{v}_i^{(\ell)})$ costs $\mathcal{O}(NM)$, due to the diagonal structure of the right multiplicand.
Therefore, computing $\pmb{C}_i$ via \eqref{eq:Ci} would necessitate $\mathcal{O}(LNM)$, hence  {$T_{\pmb{C}} = \mathcal{O}(L^2NMM_t)$}.
The complexity of solving $(\mathcal{P}_2)$ in \eqref{eq:main-subproblem-3} depends on the integrated solver, but a common well-known approach in optimization theory to solve an \ac{SDP} is via interior-point methods \cite{boyd2004convex}, e.g. \textit{primal-dual path-following algorithm} \cite{doi:10.1137/0806020}. With such an approach, and following \cite{5447068}, we have that 
\begin{equation}
\label{eq:T_P_2}
	T_{\mathcal{P}_2}
	=
	\mathcal{O}\left(\max \{LM_t,3\}^4 \sqrt{LM_t} \log (1 / \epsilon_2)\right),
\end{equation}
where $\epsilon_2 > 0$ is the solution accuracy. Combining all the complexities, we get
\begin{equation}
	\label{eq:final-complexity-of-alg2}
\begin{split}
	T_{\rm{Alg},2}
	&=
	\mathcal{O}\left(\max \{LM_t,3\}^4 \sqrt{LM_t}N_{\rm{iter}} \log (1 / \epsilon_2)\right)
	\\&+
	\mathcal{O}
	\left(
	LM_t(ML+N)^2 I N_{\rm{iter}}\log (1 / \epsilon_1) \right)
	+
	\\&+
	\mathcal{O} \left((L^2 M_t MN I + L^2NMM_t+L^2MM_t^2) N_{\rm{iter}}
	\right) 
	\\& 
		+\mathcal{O}
	\left(
	L^3M_t^3N_{\rm{iter}}
	\right).
\end{split}
\end{equation}

\section{Simulation Results}
\label{sec:simulation-results}
\begin{figure}[t]
	\centering
\includegraphics[width=1\linewidth]{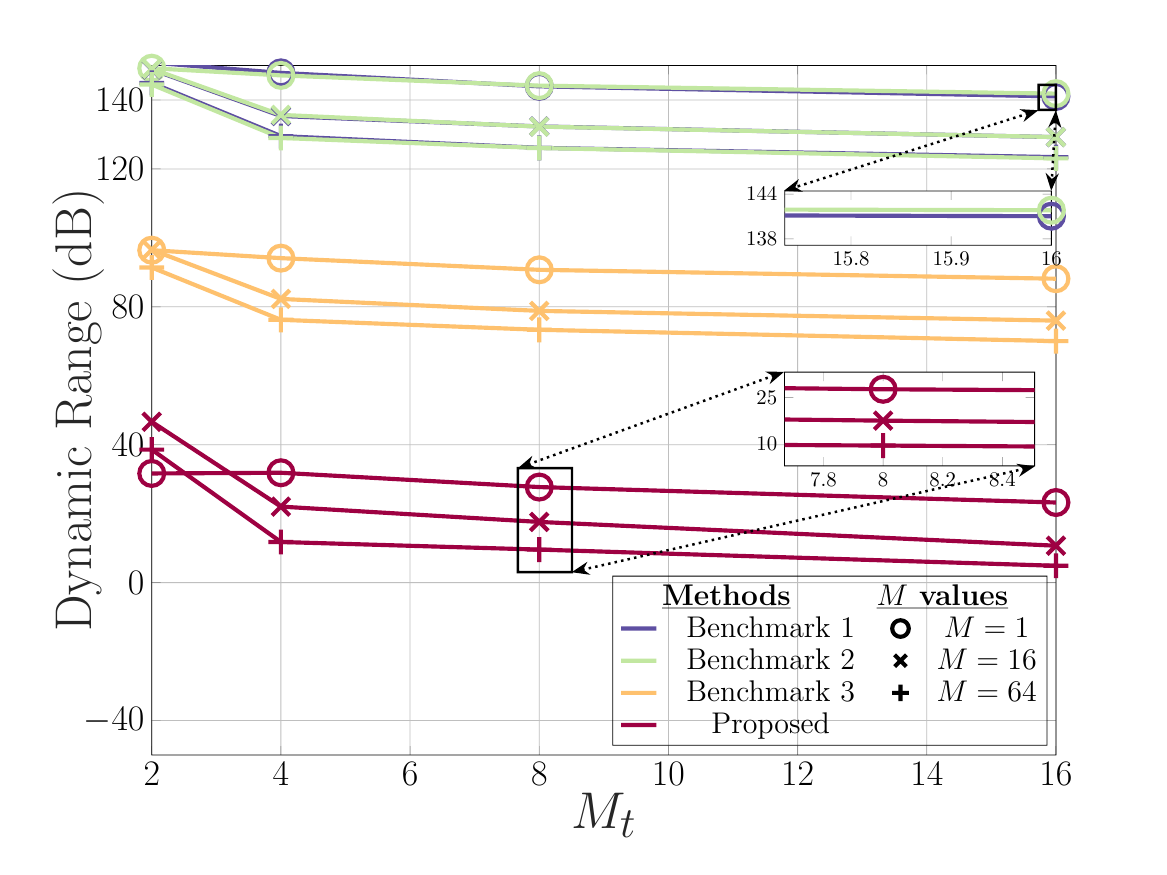}
\input{Actions/fig3.tex}
	\label{fig:DR_vs_Mt}
\end{figure}

\begin{figure}[t]
	\centering
\includegraphics[width=1\linewidth]{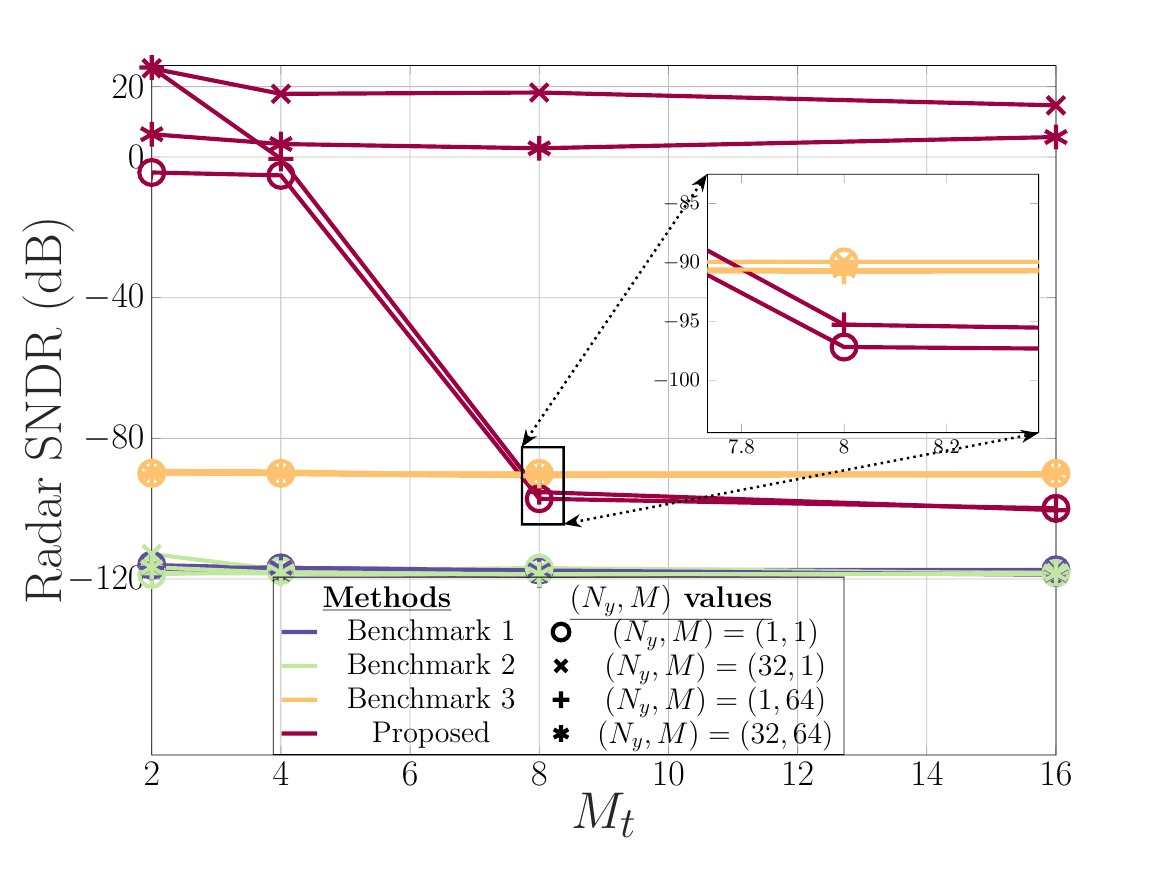}
\input{Actions/fig4.tex}
	\label{fig:SNDR_vs_Mt}
\end{figure}

\begin{figure}[t]
	\centering
\includegraphics[width=1\linewidth]{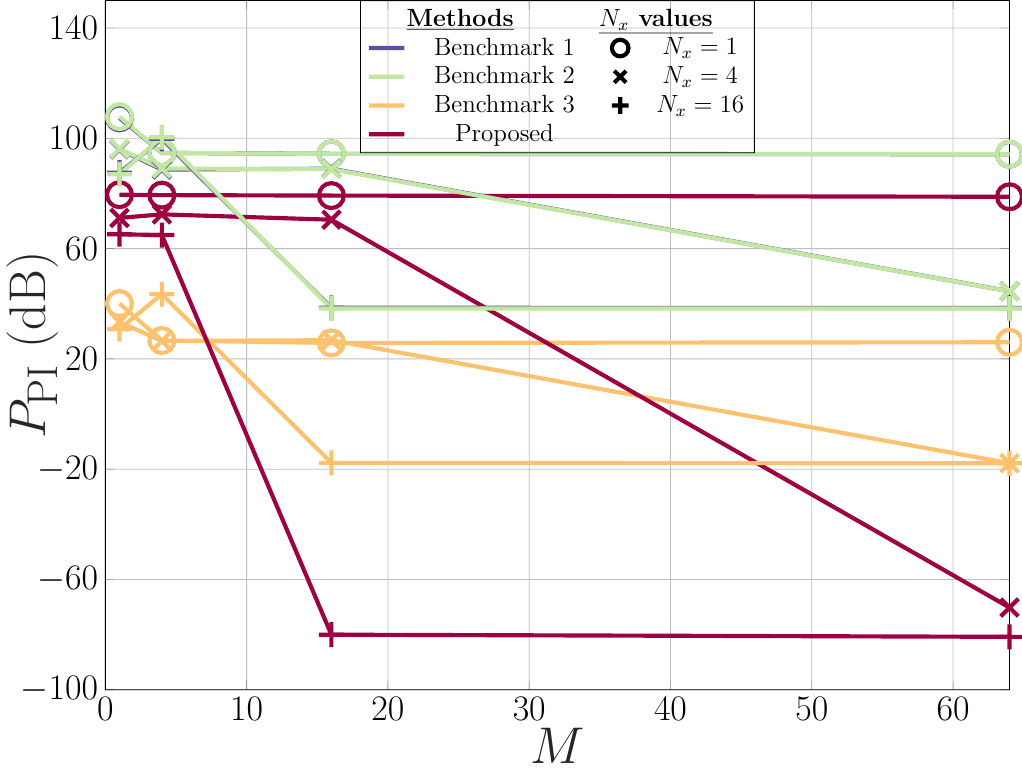}
	\input{Actions/fig5.tex}
	\label{fig:PI_vs_M}
\end{figure}

\begin{figure}[t]
	\centering
\includegraphics[width=1\linewidth]{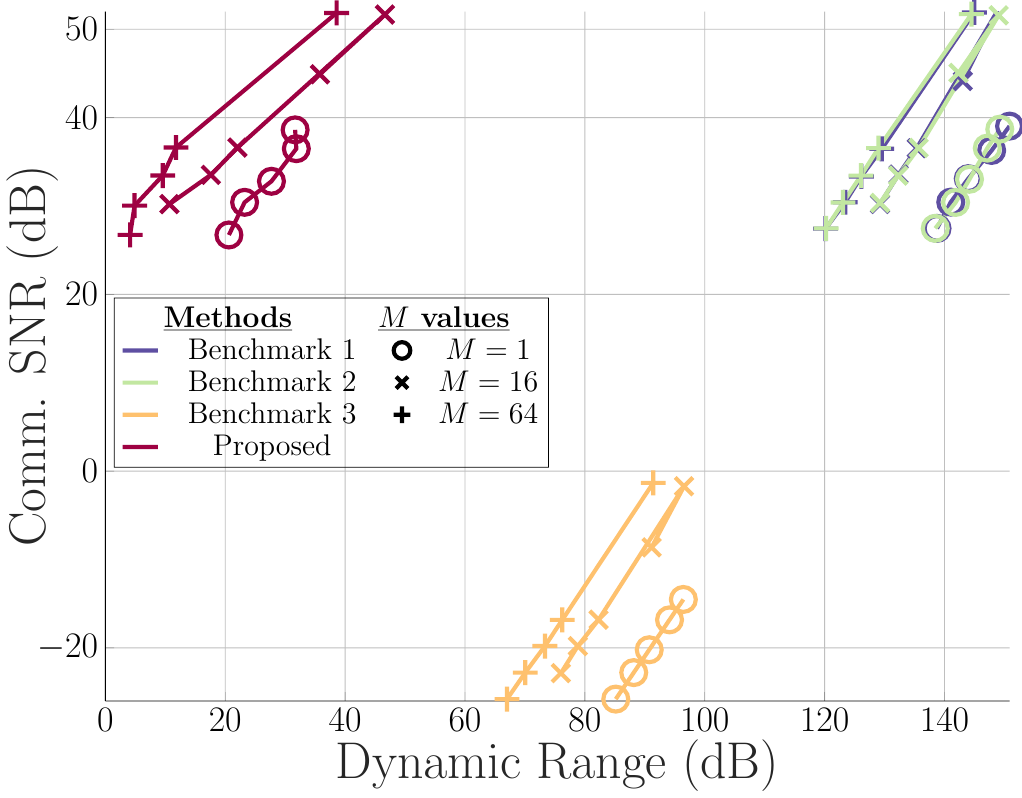}
\input{Actions/fig6.tex}
	\label{fig:CommSNR_vs_DR}
\end{figure}

\begin{figure}[t]
	\centering
\includegraphics[width=1\linewidth]{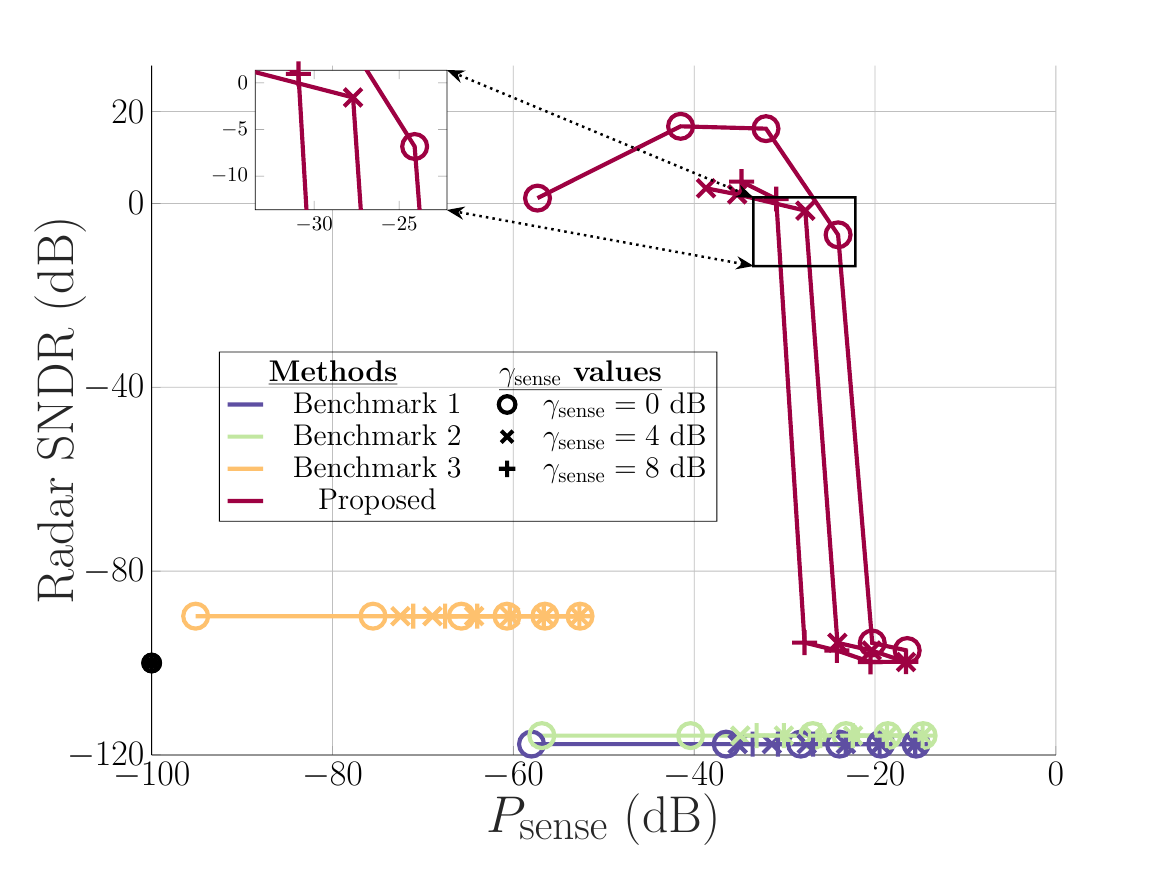}
\input{Actions/fig7}
	\label{fig:RadarSNR_vs_Psense}
\end{figure}

\begin{figure}[t]
	\centering
\includegraphics[width=1\linewidth]{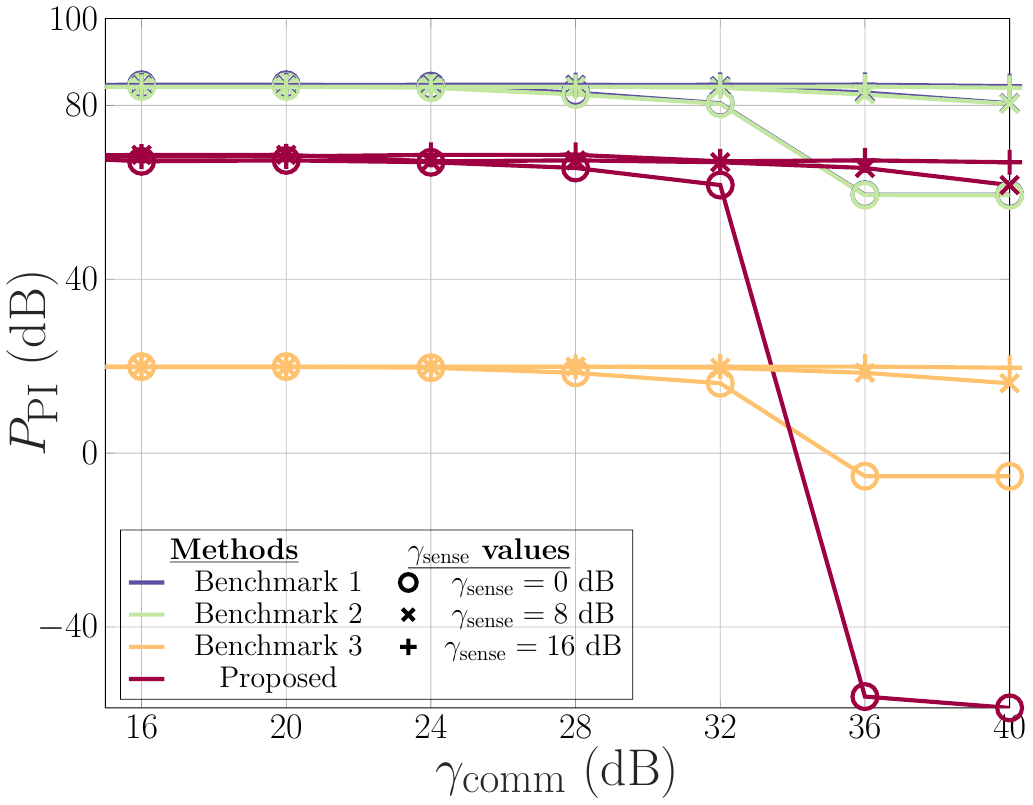}
\input{Actions/fig8}
	\label{fig:PI_vs_gammaComm}
\end{figure}

\begin{figure}[t]
	\centering
\includegraphics[width=1\linewidth]{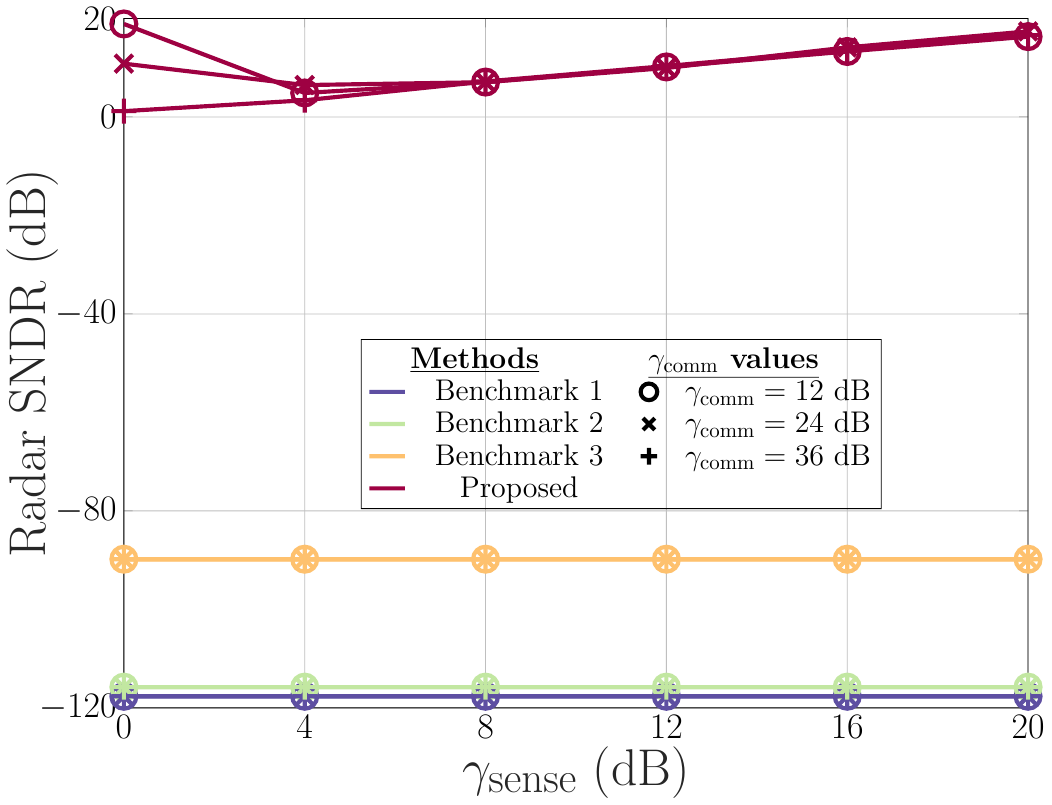}
\input{Actions/fig9}
\label{fig:RadarSNR_vs_gammaSense}
\end{figure}

In this section, we present our simulation findings. 
Before we analyze and discuss our results, we mention the simulation parameters and benchmarks used in simulations.
\subsection{Parameter Setup}
\label{subsec:parameter-setup}
Unless otherwise stated, we use the simulation parameters depicted in Table \ref{tab:table1}.
Monte Carlo type simulations is conducted in order to analyze the efficacy of the proposed method.
In typical radar applications \cite{Raghunandan2022}, as well as communications \cite{9226127}, \acp{LNA} are specially designed to receive weak signals by boosting the incoming signal level.
The cell radius is set to $500$ meters \cite{8340227}.
Moreover, the carrier frequency is $28\GHz$ and the thermal noise is $-174\dBmpHz$.
Furthermore, the path-loss exponent is set to reflect a free space setting, i.e. $2$.
The antennas follow a \ac{ULA} structure with spacing of half a wavelength, i.e. $\frac{\lambda}{2}$.
At the \ac{PR}, we have set the \ac{LNA} gain to $40\dB$ gain, following \cite{10353001} and the receive antenna gain at the \ac{PR} is $25\dBi$.
At the \ac{RIS}, we have set the reflecting \ac{RIS} coefficient to unity, i.e. $A = 1$. Moreover, the single \ac{RIS} element size is $d_x = d_y = 0.4\lambda$ \cite{10103813}.
At the \ac{BS}, the transmit power is $40\dBm$, and the transmit antenna gain is $25\dBi$ \cite{8207426}.
At the user, the receive antenna gain is $12\dBi$ \cite{8207426}.
The channel generated between nodes is Rayleigh. Also, we have set $Q = 0$.

\begin{table}[!t]
\caption{Simulation Parameters\label{tab:table1}}
\centering
{
\begin{tabular}{|c||c|}
\hline
\textbf{Parameter} & \textbf{Value}\\
\hline
Number of cells & $1$ \\
\hline
Cell radius & $500\meters$ \cite{8340227} \\
\hline
Carrier frequency $(f_c)$& $28\GHz$  \\
\hline
Thermal noise& $-174\dBmpHz$ \\
\hline 
Path-loss exponent & $2$ (free space) \\
\hline
Antenna spacing & $\frac{\lambda}{2}$ \\
\hline 
Reflecting \ac{RIS} coefficient $(A)$ & $1$ \\
\hline
Single \ac{RIS} element size ($d_x,d_y$) & $0.4\lambda$ \cite{10103813} \\
\hline
Transmit power by \ac{BS} & $40\dBm$ \cite{8207426}  \\
\hline
Transmit antenna gain at \ac{BS} & $25\dBi$ \cite{8207426}  \\
\hline
Receive antenna user gain & $12\dBi$ \cite{8207426}  \\
\hline
Receive antenna \ac{PR} gain & $25\dBi$ \cite{8207426}  \\
\hline
\ac{LNA} gain at \ac{PR} & $40\dB$ \cite{10353001}  \\
\hline
Antenna geometry & \ac{ULA} \\
\hline
Channel conditions & \makecell{Rayleigh}\cite{8207426}\\
\hline
\end{tabular}
}
\end{table}

\subsection{Benchmark Schemes}
Throughout simulations, we compare our proposed method with the following three schemes: 
\begin{itemize}
	\item The random phase shift design where arbitrary values in the range $[0,2\pi]$ are considered. In figure legends, this benchmark is denoted by \textbf{Benchmark 1}. 
	Note that this benchmark is widely used within the \ac{RIS} literature \cite{9133142}, \cite{10301482}, \cite{9614196}.
	
	\item The equal phase shift design where we fix the phase shifts of the \ac{RIS} to be all equal to one another. In figure legends, this benchmark is denoted by \textbf{Benchmark 2}. 
	Note that this benchmark is also common within the \ac{RIS} \cite{10301482}, \cite{10136735} and \ac{STAR}-\ac{RIS} \cite{10297571} literature.
	
	\item  This is a benchmark with no RIS deployment, and is denoted as \textbf{Benchmark 3} within the legends of the simulation figures.  Only direct links are considered between 
	\ac{BS}-user, 
	\ac{BS}-target, 
	\ac{BS}-\ac{PR}, and
	target-\ac{PR}.
	Note that the no-\ac{RIS} baseline is also a popular one within the \ac{RIS}
	\cite{10225701}, \cite{10280714}, \cite{9852389} literature.
	\textit{A very important remark to highlight is the fact that \ac{RIS} dimensions ($N_x,N_y$) appear on the legends of the no-\ac{RIS} benchmark. This is due to the fact that the power consumed by our method is then passed to each other benchmark, including this one. Therefore, we have kept the \ac{RIS} dimensions to have a fair assessment within simulations. In other words, all benchmarks operate at the same consumed power, i.e. $\operatorname{tr}\left(\boldsymbol{R}_{ss}\right)$.}
\end{itemize}

\subsection{Simulation Results}

\paragraph{Dynamic range behavior} 
In Fig. \ref{fig:DR_vs_Mt}, 
we fix the number of receiving antennas at the user to $M_r = 32$. 
In addition, the \ac{RIS} size is $N_x=N_y=32$.
The number of time samples is $L = 5$ samples, the communication noise variance was set to $\sigma_c^2 = - 80\dBm$, the radar noise variance was fixed at $\sigma_r^2 = - 80\dBm$.
The power budget for the optimization method is $P_B = 0 \dB$.
In addition, we set $\gamma_{\rm{comm}} = 0 \dB$ and $\gamma_{\rm{sense}} = 0 \dB$.
We run the proposed method and average over Monte Carlo trials.
The simulations reveal that increasing the number of transmit antennas, $M_t$ can contribute to a decrease of the dynamic range at the \ac{PR} for all benchmarks, as well as the proposed method.
Generally speaking, and in the high $M_t$ regime, doubling the number of antennas decreases the dynamic range by roughly $5\dB$. 
In addition, the number of antennas at the \ac{PR} can aid in analog space-time beamforming to further lower down the dynamic range.
For instance, at $M_t = 8$ antennas, the proposed method shows almost $10\dB$ reduction between when going from $M=1$ to $M=16$, in addition to another $10\dB$ reduction when increasing $M$ from $16$ to $64$.
Moreover, the proposed method can enormously reduce the dynamic range by factors of order of $50\dB$ as compared to a no-\ac{RIS} setting (i.e. \textbf{Benchmark 3}).
This can be explained by the accommodation of the \ac{PI} within the optimization framework and accounting for its minimization rather than dealing with it in a separate stage. 
We also observe that \textbf{Benchmark 1} and \textbf{Benchmark 2} reveal similar performance.
\paragraph{Radar \ac{SNDR} stability}
In Fig. \ref{fig:SNDR_vs_Mt}, we set the values of $M_r,N_x,L,\sigma_c^2,\sigma_r^2,P_B,\gamma_{\rm{comm}}, \gamma_{\rm{sense}}$ are the same as those in Fig. \ref{fig:DR_vs_Mt}.
This simulation shows that increasing the number of transmit antennas, $M_t$ alone cannot guarantee a consistent radar \ac{SNDR}, and increasing the \ac{RIS} size can help overcome this issue through the proposed method.
For example, when the number of \ac{RIS} elements along the y-axis is set to only $N_y = 1$, then we can see that then radar \ac{SNDR} drops after $8$ \ac{BS} antennas (for both $M = 1$ and $M = 64$ cases) to below the case of no-\ac{RIS} indicating inefficient use of available power resources.
Meanwhile, when $N_y=32$ elements is used, the radar \ac{SNDR} can be guaranteed to remain high.
This is because more transmit antennas can contribute to deteriorate the radar \ac{SNDR} by unwillingly injecting more \ac{PI} into the model. 
The \ac{RIS} can overcome this leakage and help maintain a high radar \ac{SNDR}.

\paragraph{\ac{PI} suppression}
%
In Fig. \ref{fig:PI_vs_M}, we set the values of $N_x,L,\sigma_c^2,\sigma_r^2,P_B,\gamma_{\rm{comm}}, \gamma_{\rm{sense}}$ are the same as those in Fig. \ref{fig:DR_vs_Mt}.
We also have $M_t = 32$ antennas, $M_r = 1$ antenna and $N_y = 32$.
This simulation shows that the \ac{PR} alone may not be fully helpful in maintaining a low \ac{PI} power level.
For instance, the \ac{PI} power level with the proposed method remains around $80\dB$ for any $M \leq 64$ when $N_x = 1$.
Upon increasing the \ac{RIS} size to just $N_x = 4$, the \ac{PI} power drops at $M = 64$ to a level beyond $-60\dB$, and, and at $M = 16$ for $N_x = 16$ also to a level beyond $-60\dB$.
Also once can see that both benchmarks, \textbf{Benchmark 1} and \textbf{Benchmark 2} perform equally the same, and that the proposed method can bring the \ac{PI} power level down beyond all the three benchmarks by $50\dB$.
This is because more \ac{RIS} elements can provide extra degrees-of-freedom to aid the \ac{PR} in its analog space-time beamforming task to further suppress the \ac{PI} via the proposed optimization method.

\paragraph{Communication \ac{SNR} \& dynamic range tradeoff}
In Fig. \ref{fig:CommSNR_vs_DR}, we set the values of $M_r,N_x,N_y,L,\sigma_c^2,\sigma_r^2,P_B,\gamma_{\rm{comm}}, \gamma_{\rm{sense}}$ the same as those in Fig. \ref{fig:DR_vs_Mt}.
This simulation shows that the proposed method can achieve high communication \ac{SNR} rates at a much lower dynamic range than the other benchmarking techniques.
For example, if one targets a communication \ac{SNR} of $30\dB$, then the proposed method can achieve this communication \ac{SNR} with about $23\dB$ dynamic range with $M = 1$ antenna at the \ac{PR}, whereby requiring more than $100\dB$ for both \textbf{Benchmark 1} and \textbf{Benchmark 2}. 
In addition, \textbf{Benchmark 3} can not achieve such a communication \ac{SNR} of $30\dB$ due to the un-involvement of a \ac{RIS} in that case.
A counter-intuitive point to highlight is that increasing $M$ can leave more room to enhance the communication \ac{SNR}. Indeed, for a fixed dynamic range, increasing $M$ can allow us to achieve a higher \ac{SNR}. 
For example, a dynamic range of $20\dB$ can only achieve a $26\dB$ communication \ac{SNR} with $M=1$, whereas the same dynamic range can achieve an \ac{SNR} of $35\dB$ for $M = 16$ and can even go beyond $40\dB$ for $M = 64$ antennas.

\paragraph{Radar \ac{SNDR} behavior}
In Fig. \ref{fig:RadarSNR_vs_Psense}, we set $M_t = 2$ antennas, $M_r = 1$ antennas, $N_x = N_y = 8$ elements, $M = 24$ antennas and $L = 5$ samples.
The value of $\sigma_c^2$ is fixed to $-100\dBm$ and $\sigma_r^2$ is set to $-110 \dBm$.
In addition, we set $\gamma_{\rm{comm}} = 36\dB$.
This simulation studies the behavior of the radar \ac{SNDR} as a function of achieved $P_{\rm{sense}}$ for different values of $\gamma_{\rm{sense}}$. 
First, notice that the benchmarks reveal very low radar \ac{SNDR} values, i.e. $-90\dB$ for \textbf{Benchmark 3} (no-\ac{RIS}) and an even lower radar \ac{SNDR} for \textbf{Benchmark 1} and \textbf{Benchmark 2}, but a larger achieved $P_{\rm{sense}}$ due to the additional (however un-optimized) paths provided by the \ac{RIS}. 
These low \ac{SNDR} values are due to the dominating \ac{PI} power level.
To restore the radar \ac{SNDR} to high values in order to achieve high target detection performance, our proposed method can reach \ac{SNDR} values of $-5\dB$ at ranges of $P_{\rm{sense}}$ within $[-32,-24]\dB$.
In addition, for a fixed radar \ac{SNDR}, $\gamma_{\rm{sense}}$ not only provides a clear control of radar \ac{SNDR}, but also allows us to achieve \ac{SNDR} targets for lower $P_{\rm{sense}}$ values. 
For example, to reach a radar \ac{SNDR} value of $-5\dB$, one would require $-24 \dB$ of $P_{\rm{sense}}$ with a $\gamma_{\rm{sense}} = 0\dB$, whereas a $-28\dB$ for $\gamma_{\rm{sense}} = 4\dB$ and $-32\dB$ for $\gamma_{\rm{sense}} = 8\dB$. 
In essence, an increase of $\gamma_{\rm{sense}}$ allows for radar sensing operations to take place at lower sensing power.

\paragraph{Prioritizing Communications can help mitigate \ac{PI}}
In Fig. \ref{fig:PI_vs_gammaComm}, we fix $P_B = 0 \dB$. Moreover, the values of $M_t$, $M_r$, $N_x$, $N_y$, $M$, $L$, $\sigma_c^2$, and $\sigma_r^2$ are the same as those in Fig. \ref{fig:RadarSNR_vs_Psense}.
This simulation studies the influence of $\gamma_{\rm{comm}}$ on the \ac{PI} power level.
We note that for all benchmarks, and the proposed method, when $\gamma_{\rm{comm}}$ dominates $\gamma_{\rm{sense}}$ (i.e. when communications is prioritized over sensing), the \ac{PI} power level experiences an abrupt decrease, with the most significant drop observed in the proposed approach.
 For example, when $\gamma_{\rm{comm}} = 40\dB$ and $\gamma_{\rm{sense}} = 0\dB$, the \ac{PI} power level corresponding to the proposed method drops to below $-40\dB$, while remaining over $60\dB$ when $\gamma_{\rm{sense}} \geq 8\dB$, which is still $15 \dB$ lower than benchmarks 1 and 2.
 This means that communication tasks in the \ac{ISAC} context can help in mitigating the \ac{PI} power level.

\paragraph{Influence of $\gamma_{\rm{sense}}$ and $\gamma_{\rm{comm}}$ on radar \ac{SNDR}}
In Fig. \ref{fig:RadarSNR_vs_gammaSense}, we fix $\sigma_r^2 = -85\dBm$. Moreover, the values of $P_B$, $M_t$, $M_r$, $N_x$, $N_y$, $M$, $L$, $\sigma_c^2$, and the same as those in Fig. \ref{fig:PI_vs_gammaComm}.
This simulation studies the joint influence of $\gamma_{\rm{sense}}$ and $\gamma_{\rm{comm}}$ on the radar \ac{SNDR}.
First, we note that $\gamma_{\rm{sense}}$ controls the radar \ac{SNDR} as expected. In particular, for the proposed method, the radar \ac{SNDR} increases with $\gamma_{\rm{sense}}$.
In addition, increasing $\gamma_{\rm{comm}}$ lowers the radar \ac{SNDR} to settle at the boundary specified by $\gamma_{\rm{sense}}$. 
Meanwhile, the radar \ac{SNDR} levels of all benchmarks are always constant in this case and are below $-80\dB$.

\section{Open Challenges}
\label{sec:open-challenges}
Developing a \ac{RIS}-aided bi-static \ac{ISAC} system presents challenges across multiple aspects. 
This paper proposes one technique for mitigating \ac{PI} by directly suppressing it while satisfying relevant \ac{ISAC} \acp{KPI}; nevertheless, alternative solutions tailored to different scenarios are also feasible.
Implementing a bi-static \ac{ISAC} system presents many open challenges, where we summarize a few below.
\subsection{Near Field \ac{ISAC} with \ac{PI} mitigation}
\label{subsec:pi-nearfield} 
\acp{RIS} have the potential to enable \ac{ISAC} applications and in some cases enhance \ac{ISAC} performance \cite{10254508}.
As far as \ac{PI} mitigation is concerned, an open future challenge worth investigating is the potential to exploit wavefront curvature in geometric near-field conditions in order to further enhance communication \ac{SNR} and radar \ac{SNDR}, whilst maintaining a very low \ac{PI} power level in near-field conditions.
As the Fraunhofer distance is enlarged, the wavefronts within this region are spherical, which breaks the planar wavefront assumption.
It is worth noting that \ac{RIS} has already been explored for near-field sensing and beamforming \cite{9941256} and localization \cite{10017173}, but remains un-explored in terms of \ac{PI} mitigation for \ac{ISAC} bi-static sensing.
It is important to note that in near-field conditions, beamforming can concentrate signals around a specific location, namely \textit{beamfocusing}, which has the full capacity of better improving the sensing and communications performance.
Enabling this would entail redesigning the statistical transmit covariance and the space-time analog beamformers for improved \ac{PI} mitigation.

\subsection{Scaling up more \acp{RIS} and \acp{PR}}
\label{subsec:scaling-up-more-RIS-PR}
While multiple \acp{RIS} and \acp{PR} can provide additional spatial diversity, which can then enable the enhancement of the sensing performance, however more \acp{PI} will arise by construction of the model.
This, in turn, opens an additional challenge and the need to jointly suppress the multiple \acp{PI} that arise.
Another challenge that comes with scaling more \acp{RIS} and \acp{PR} within the scene backhauling capacity connected with the \ac{CPU}.
In particular, if limited capacity backhaul links are installed, then a bottleneck can arise, hence bi-directional communication between the \ac{CPU} and the \ac{PR}, \ac{RIS} and \ac{BS} can be an additional challenge when incorporating additional \acp{RIS} and \acp{PR}.
This can include traffic to prioritize, in addition to the quantities to be exchanged, given throughput limitations. 
More specifically, once the \ac{CPU} computes the statistical covariance matrix for the \ac{BS}, along with the \ac{RIS} phase shifts and the space-time beamformer, a fundamental question from backhauling perspective is, \textit{"Should the entire quantities be forwarded to the intended units, or should a compressed version of these quantities be sent?"}
Another question we can ask when scaling up \acp{PR} is that, \textit{"Can some of the processing be performed in a distributed manner ? Said differently, can a distributed optimization algorithm be tailored to split the processing and finally reducing traffic exchange to avoid possible bottlenecks ?"} 
This would require decomposing the formulated optimization problem in a distributed way requiring minimal input exchanges between the underlying subproblems.

\subsection{\ac{ADC} Optimization}
\label{subsec:adc-opt}
An interesting line of research well-suited in this challenge is the few-bit (including $1$-bit) \ac{ADC} design.
The goal is to design the \ac{ADC} with smaller dynamic range that can accommodate signals with large power discrepancies \cite{8571272}, \cite{9695370}, \cite{8606934}.	
For instance, the work in \cite{9054740} proposes a generalized approximate message passing algorithm for low-resolution \ac{ADC} that observed a $15\dB$ dynamic range reduction.
An open challenge would be to design algorithms tailored for low-resolution (few-bit or/and $1$-bit) \acp{ADC} to be able to extract the meaningful sensing data despite the presence of large \ac{PI} power levels. 
To realize this, one common approach would be to introduce a $b-$bit quantization function, say $\mathcal{Q}_b\left(.\right)$, to capture the quantization effects of the \ac{ADC}.
The received signal at the \ac{PR}, and after analog beamforming, would be inputted to the \ac{ADC}, i.e. $\mathcal{Q}_b\left(\pmb{w}^H \pmb{y}_t \right)$.
The goal now would be to sense the target by observing $\mathcal{Q}_b\left(\pmb{w}^H \pmb{y}_t \right)$ instead of its continuous version, i.e. $\pmb{w}^H \pmb{y}_t$.
Another approach would be to have an \ac{ADC} per receive chain, which means that the \ac{PR} would have to integrate $M$ \acp{ADC} instead of just one, in the case of analog beamforming, which would add more \ac{PR} processing complexity.

\subsection{Imperfect channel state information}
\label{subsec:imperfect-csi-challenge}
A more practical approach is to mitigate \ac{PI} while carrying out \ac{ISAC} tasks, even in the presence of imperfect \ac{CSI}.
The primary factors leading to imperfect \ac{CSI} are inaccurate channel estimation and hardware imperfections \cite{10018908}. In fact, ignoring these issues could significantly degrade overall system performance if they are not effectively addressed.
To consider imperfect \ac{CSI}, we may assume that the \ac{BS} has access to \ac{CSI} information in the form
\begin{equation}
\label{eq:csi-Hk}
\begin{split}
\boldsymbol{H}_{\mathrm{k}}&=\widehat{\boldsymbol{H}}_{\mathrm{k}}+\Delta \boldsymbol{H}_{\mathrm{k}},
\end{split}
\end{equation}
where $\widehat{\boldsymbol{H}}_{\mathrm{k}}$ is the estimated \ac{CSI} and $\Delta \boldsymbol{H}_{\mathrm{k}}$ stands for the \ac{CSI} errors.
A similar approach could be applied to the channels $\boldsymbol{H}_{R k}$ and $\boldsymbol{H}_{\mathrm{cR}}$, as 
\begin{equation}
\begin{split}
\boldsymbol{H}_{\mathrm{cR}}&=\widehat{\boldsymbol{H}}_{\mathrm{cR}}+\Delta \boldsymbol{H}_{\mathrm{cR}},\\
\label{eq:csi-HcR}
\boldsymbol{H}_{\mathrm{Rk}}&=\widehat{\boldsymbol{H}}_{\mathrm{Rk}}+\Delta \boldsymbol{H}_{\mathrm{Rk}},
\end{split}
\end{equation}
where $\widehat{\boldsymbol{H}}_{\mathrm{cR}},\widehat{\boldsymbol{H}}_{\mathrm{Rk}}$ represent the estimated \ac{CSI} from \ac{BS} to \ac{RIS} and \ac{RIS} to user, respectively. In addition, $\Delta {\boldsymbol{H}}_{\mathrm{cR}},\Delta {\boldsymbol{H}}_{\mathrm{Rk}}$ represent their corresponding channel errors, respectively.
Error modeling depends on whether a stochastic or deterministic  approach is taken. In the former, the errors are usually assumed to be normally distributed.
The latter, i.e. deterministic and norm bounded, corresponds to worst case optimization. 
One way to model this is to restrict the errors to live in a certain space, i.e. $\left\|\boldsymbol{E}_{\mathrm{k}}^{\frac{1}{2}} \Delta \boldsymbol{H}_{\mathrm{k}}\right\| \leq 1,$ where $\boldsymbol{E}_{\mathrm{k}}$ controls the ellipsoidal shape of the \ac{CSI} perturbation set for the direct channel. A similar restriction could be applied onto $\Delta \boldsymbol{H}_{\mathrm{Rk}}$ and $\Delta \boldsymbol{H}_{\mathrm{cR}}$.
For sensing, we could also model the sensing channels, i.e. the one-way, two-way, three-way and four-way channels, as done in \eqref{eq:csi-Hk}, \eqref{eq:csi-HcR}. 
An alternative, in case of strong \ac{LoS} with the target, would be to assume uncertainty on some location information of the target itself, i.e. assume the target is at \ac{AoA} $\theta_{\rm{t}}$ from the \ac{PR} and at \ac{AoD} $\phi_{\rm{t}}$ from the \ac{BS}, but the \ac{BS} and target have partial knowledge of this. This means that the assumed channels would be
$\widehat{\boldsymbol{g}}_{\mathrm{t}} = \pmb{a}_M(\tilde{\theta}_{\rm{t}})$, where $\tilde{\theta}_{\rm{t}} \in \left[{\theta}_{\rm{t}} - \frac{\Delta \theta}{2};{\theta}_{\rm{t}} + \frac{\Delta \theta}{2} \right]$ and $\widehat{\boldsymbol{h}}_{\mathrm{t}} = \pmb{a}_{M_t}(\tilde{\phi}_{\rm{t}})$, where $\tilde{\phi}_{\rm{t}} \in \left[{\phi}_{\rm{t}} - \frac{\Delta \phi}{2};{\phi}_{\rm{t}} + \frac{\Delta \phi}{2} \right]$. This modeling captures the case of any angle uncertainty on the location of the target. Indeed, the case of $\Delta \theta = \Delta \phi = 0$ corresponds to the case of perfect knowledge case. The same modeling could be applied onto the second, third and fourth order channels.
The remaining challenge is to formulate a suitable optimization problem that can still be able to mitigate the \ac{PI} while satisfying the minimal acceptable sensing and communication performance in the presence of all the \ac{CSI} errors.

\section{Conclusion}
\label{sec:conclusion}

This paper presented a novel framework for optimizing the dynamic range at a \ac{PR} installed within a \ac{RIS}-aided \ac{ISAC} bistatic system.
Due to \ac{RIS} presence, additional interfering paths are present in the model, which we call \ac{PI}.
Moreover, we have introduced a dynamic range optimization approach that leverages space-time analog beamforming, in conjunction with \ac{RIS} phase shift optimization to mitigate path interference \ac{PI}, while maintaining both communication and sensing performance. The proposed \ac{BCCD} algorithm effectively solves the resulting non-convex optimization problem, leading to significant improvements in system performance metrics such as \ac{SNDR} and communication \ac{SNR}.
We have also analyzed the complexity of the proposed methods.
This underscores the potential \ac{ISAC} applications to exist with low dynamic range, particularly in the presence of multiple sensing reflectors.

\appendices
\label{sec:appendix}
\section{Problem Equivalence}
\label{app:problem-equivalence}
For ease of representation, we first denote $\bar{\boldsymbol{H}}_{\rm{PI}} = \bar{\boldsymbol{H}}_{\rm{DPI}} + \bar{\boldsymbol{H}}_{\rm{RPI}}$,
where $ \bar{\boldsymbol{H}}_{\rm{DPI}}= \gamma_{\rm{DPI}}
\boldsymbol{H}_{\rm{DPI}}$ and $ \bar{\boldsymbol{H}}_{\rm{RPI}}= \gamma_{\rm{RPI}}
\boldsymbol{H}_{\rm{RPI}}$
To this end, we have the following series of steps 
\begin{equation*}
\begin{split}
	P_{\rm{PI}}
	&=
	{\pmb{w}}^H\boldsymbol{A}_c(\phi) \pmb{R}_{ss} 
	\boldsymbol{A}_c^H (\phi) 
	{\pmb{w}} \\
	&\myeqa \sum_i \lambda_i \vert {\pmb{w}}^H\boldsymbol{A}_c(\phi) \pmb{v}_i \vert^2\\
	&\myeqb \sum_i \lambda_i \vert {\pmb{w}}^H (\boldsymbol{I}_L \otimes \bar{\boldsymbol{H}}_{\rm{PI}}) \pmb{v}_i \vert^2\\
	&\myeqc \sum_i \lambda_i \vert {\pmb{w}}^H (\boldsymbol{I}_L \otimes \bar{\boldsymbol{H}}_{\rm{DPI}}
+
\boldsymbol{I}_L \otimes\bar{\boldsymbol{H}}_{\rm{RPI}}) \pmb{v}_i \vert^2\\
	&\myeqd \sum_i \lambda_i \vert {\pmb{w}}^H (\boldsymbol{I}_L \otimes \bar{\boldsymbol{H}}_{\rm{DPI}}) \pmb{v}_i+ {\pmb{w}}^H (\boldsymbol{I}_L \otimes\bar{\boldsymbol{H}}_{\rm{RPI}}) \pmb{v}_i \vert^2\\
&\myeqe \sum_i \lambda_i \vert {\pmb{w}}^H (\boldsymbol{I}_L \otimes \bar{\boldsymbol{H}}_{\rm{DPI}}) \pmb{v}_i +  {\pmb{w}}^H (\boldsymbol{I}_L \otimes \boldsymbol{G}_{\rm{r R}}^{H} \boldsymbol{\Phi} \bar{\boldsymbol{H}}_{\rm{c R}}) \pmb{v}_i \vert^2\\
&\myeqf \sum_i \vert \pmb{w}^H \pmb{b}_i +  {\pmb{w}}^H \pmb{C}_i\pmb{\phi}	 \vert^2,\\
\end{split}
\end{equation*}
where in step $(a)$, we have decomposed $\pmb{R}_{ss}$ in terms of its eigenvalue decomposition, i.e. $\lambda_i$ and $\pmb{v}_i$ represent the $i^{th}$ eigenvalue and eigenvector of $\pmb{R}_{ss}$, respectively.
In step $(b)$, we have used equation \eqref{eq:Ac}.
In step $(c)$, we have used equation \eqref{eq:decompose-PI}.
In step $(d)$, we have distributed the inner-product.
In step $(e)$, we have used equation \eqref{eq:RPI}.
Finally, in step $(f)$, we have used the diagonal commutative property, i.e. $\pmb{\Phi} \pmb{a} = \operatorname{diag}(\pmb{a})\boldsymbol{\phi}$, which holds true for any vector $\pmb{a}$.
The expressions of $\pmb{b}_i$ and $\pmb{C}_i$ are given in equations \eqref{eq:bi} and \eqref{eq:Ci}, respectively.  


\bibliographystyle{IEEEtran}
\bibliography{refs}

\vfill

\end{document}